%% file: main.tex
\documentclass[10pt, conference,letterpaper]{IEEEtran}
\makeatletter
\def\ps@headings{%
\def\@oddhead{\mbox{}\scriptsize\rightmark \hfil 
}%
\def\@evenhead{\scriptsize
\hfil \leftmark\mbox{}}%
\def\@oddfoot{}%
\def\@evenfoot{}}
\usepackage{booktabs}
\makeatother 

\usepackage{colortbl}
\usepackage{xcolor}
\usepackage{array}

\usepackage{graphicx}
\usepackage{epstopdf}
\usepackage{cite} 
\usepackage{times}
\usepackage{dsfont} 
\usepackage{cite}
\usepackage{times}
\usepackage{pifont}
\usepackage{multirow}   
\usepackage{tikz}
\usetikzlibrary{shapes.geometric,calc}
   
\usepackage{amsthm}
\usepackage{graphicx}
\usepackage{subfigure}
\usepackage{color}
\usepackage{ifpdf}
\usepackage{epsfig}
\usepackage{latexsym}
\usepackage{amsfonts}
\usepackage{amssymb}
\usepackage{paralist}
\usepackage{comment}
\usepackage{xspace}
\usepackage{mathrsfs}
\usepackage{amssymb}
\usepackage{setspace}
\usepackage{color}
\usepackage[small]{caption}

\usepackage{subeqnarray}
\usepackage{amssymb}
\usepackage{url,epsfig,array}
\usepackage{leftidx}
\usepackage{amsmath}
\usepackage[T1]{fontenc}
\usepackage{aecompl}

\usepackage{calligra}
\usepackage{algorithm}
\usepackage{algorithmicx}
\usepackage{algpseudocode} 
\usepackage{amsmath} 
\renewcommand{\algorithmicrequire}{\textbf{Input:}}
\renewcommand{\algorithmicensure}{\textbf{Output:}}

\def\ie{\textit{i.e.}\xspace}

\def\eg{\textit{e.g.}\xspace}

\makeatletter
\renewcommand{\maketag@@@}[1]{\hbox{\m@th\normalsize\normalfont#1}}%
\makeatother

\begin{document}
  
\title{\LARGE Heroes: Lightweight Federated Learning with Neural Composition and Adaptive Local Update in Heterogeneous Edge Networks}


\author{\IEEEauthorblockN{Jiaming Yan$^{1,2}$ \ \ $^{*}$Jianchun Liu$^{1,2}$ \ \ Shilong Wang$^{1,2}$ \ \ $^{*}$Hongli Xu$^{1,2}$ \ \ Haifeng Liu$^{3}$ \ \ Jianhua Zhou$^{3}$   \\}
\IEEEauthorblockA{  
$^1$School of Computer Science and Technology, University of Science and Technology of China, China\\
$^2$Suzhou Institute for Advanced Research, University of Science and Technology of China, China\\
$^3$ Guangdong OPPO Mobile Telecommunications Corp., Ltd. Dongguan, Guangdong, China
} }

\maketitle

\begin{abstract}
\input{content/abstract.tex}

\end{abstract}
 
\begin{IEEEkeywords}
\emph{Federated Learning, Heterogeneity, Neural Composition, Local Update Frequency}.
\end{IEEEkeywords}


\section{Introduction}\label{sec:intro}
\input{content/intro.tex}

\section{Background and Motivation}\label{sec:motivation}
\input{content/motivation.tex}

\section{Proposed Framework}\label{sec:framework}
\input{content/framework.tex}

\section{Convergence Analysis}\label{sec:analysis}
\input{content/analysis.tex}

\section{Problem Formulation and Algorithm Design}\label{sec:algorithm}
\input{content/algorithm.tex}

\section{Performance Evaluation}\label{sec:evaluation}
\input{content/evaluation.tex}

\section{Related Work}\label{sec:related}
\input{content/related.tex}

\section{Conclusion}\label{sec:conclusion}
\input{content/conclusion.tex}


\bibliographystyle{IEEEtran}
\bibliography{content/refs} 


\end{document}

%% file: content/abstract.tex
Federated Learning (FL) enables distributed clients to collaboratively train models without exposing their private data.
However, it is difficult to implement efficient FL due to limited resources.
Most existing works compress the transmitted gradients or prune the global model to reduce the resource cost, but leave the compressed or pruned parameters under-optimized, which degrades the training performance.
To address this issue, the neural composition technique constructs size-adjustable models by composing low-rank tensors, allowing every parameter in the global model to learn the knowledge from all clients. 
Nevertheless, some tensors can only be optimized by a small fraction of clients, thus the global model may get insufficient training, leading to a long completion time, especially in heterogeneous edge scenarios.
To this end, we enhance the neural composition technique, enabling all parameters to be fully trained.
Further, we propose a lightweight FL framework, called Heroes, 
with enhanced neural composition and adaptive local update.
A greedy-based algorithm is designed to adaptively assign the proper tensors and local update frequencies for participating clients according to their heterogeneous capabilities and resource budgets.
Extensive experiments demonstrate that Heroes can reduce traffic consumption by about 72.05\% and provide up to 2.97$\times$ speedup compared to the baselines.

%% file: content/intro.tex
In traditional machine learning approaches, data is gathered from various sources and transmitted to a central server for model training. 
However, in edge computing (EC), where data is generated and processed at the edge clients, this centralized approach becomes infeasible due to privacy concerns \cite{liu2022enhancing}.
Federated Learning (FL) \cite{mcmahan2017communication} is an emerging paradigm that enables collaborative model training across distributed clients, bringing the power of machine learning to EC \cite{gao2019auction}. 
As the field of FL continues to evolve, it carries significant potential to advance edge computing capabilities and facilitate intelligent applications across diverse domains, such as healthcare, smart cities and autonomous vehicles\cite{kairouz2021advances}.

However, it is difficult to implement efficient FL in practical edge networks due to resource limitation \cite{liu2023finch}, which primarily revolve around constrained computation resources and scarce communication resources on edge clients, such as smartphones and Internet of Things (IoT) devices.
Firstly, edge clients often have limited computation capabilities, especially compared to high-performance devices like desktop GPUs \cite{ignatov2019ai}.
For example, the floating-point operations per second (FLOPS) of an iPhone14 is only about 5\% of that of the desktop GPU RTX3090.
Secondly, the communication resources are also scarce in edge networks \cite{liu2023yoga}.
The bandwidth of wide area networks (WANs) between the PS and edge clients is much lower (\eg, 15$\times$) than that of local area networks (LANs) within the data centers \cite{wang2021resource}.
Unfortunately, the growing complexity (\eg, parameter size or architecture) of deep neural networks (DNNs) further intensifies the consumption of both computation and communication resources. 
For instance, a standard ResNet-18 \cite{he2016deep} consists of 11.68 million parameters and requires 27.29 billion floating-point operations (FLOPs) to process a single image, making the local update process extremely slow or even infeasible on edge clients \cite{jiang2022fedmp}.
Besides, the frequent transmissions of a large number of parameters will also strain the limited network bandwidth.

Some previous works have made efforts to address the challenge of resource limitation in FL by employing various techniques, such as gradient compression \cite{li2021talk, liu2022communication} and model pruning \cite{diao2020heterofl, horvath2021fjord}. Specifically, \textit{FlexCom} \cite{li2021talk} and \textit{AdaGQ} \cite{liu2022communication} compress the transmitted gradients by sparsification or quantization to alleviate the communication overhead.
However, these approaches do not reduce the model complexity and the computation overhead is still high.
To save both the computation and communication resources, \textit{HeteroFL} \cite{diao2020heterofl} and \textit{Fjord} \cite{horvath2021fjord} propose to prune the complete model into the smaller sub-models for training.
Nevertheless, excessive parameter pruning will significantly degrade training performance.
For instance, \textit{HeteroFL} prunes 93.75\% of the parameters from the complete model to accommodate the limitations (\eg, CPU power, RAM, energy) on weak clients like smartphones, leaving only 6.25\% of the parameters to be optimized. 
As a result, the majority of parameters are unable to benefit from the local data on these weak clients, leading to poor training performance of the complete model.

To address the above issues, \textit{Flanc} \cite{mei2022resource} proposes the neural composition technique, which approximates each model weight as the product of two low-rank tensors, named neural basis and coefficient.
Concretely, the models of various complexities are constructed by composing (\ie, multiplying) the neural basis with the coefficient of different sizes.
In each round, the parameter server (PS) sends the neural basis and a proper coefficient to each client for composition and local training.
The entire neural basis is trained by all clients and its learned knowledge can be propagated to all parameters, which enables every parameter to access the full range of knowledge.
We observe that the size of low-rank tensors is inherently much smaller than that of the original model.
For example, the size of standard ResNet-18 is 42.8MB, while that of approximated tensors is only 15.3MB.
Thus, this approach can reduce both computation and communication consumption during training.

Despite resource efficiency, the neural composition technique will encounter some problems in the context of FL.
Firstly, \textit{Flanc} only aggregates the coefficients with the same shape.
As a consequence, the coefficient of a specific shape is only trained by the clients with the corresponding computation power, which may be insufficient for global model convergence.
This issue becomes more prominent when high-performance clients only constitute a small fraction of all clients in the EC system due to their expensive price. For instance, the largest coefficient is only trained by a few powerful clients.
As a result, the largest global model may struggle to converge within the given completion time, leading to poor training performance.
Secondly, the edge clients are equipped with different hardware, thus their capabilities (\eg, CPU power, bandwidth) may vary significantly \cite{li2022pyramidfl}, \ie, client heterogeneity, which poses a great impact on training efficiency.
For example, if the PS sends the neural basis and a large coefficient to a client with strong computation power but low upload bandwidth, the completion time for model updates will be prolonged, causing long delays in the aggregation step. 

In order to tackle the aforementioned challenges, we propose a lightweight FL framework, called Heroes (Lig\textbf{H}tweight F\textbf{e}derated Lea\textbf{r}ning thr\textbf{o}ugh N\textbf{e}ural Compo\textbf{s}ition).
On the one hand, we enhance the neural composition technique to aggregate the coefficients with different shapes into the largest one.
Besides, by adaptively assigning the coefficients to clients, each parameter in the global coefficient can be fully trained, ensuring global model convergence.
On the other hand, we adjust the local update frequencies for different clients to balance their completion time, so as to diminish the impact of client heterogeneity and improve the training efficiency.
Nevertheless, the neural composition and local update frequency are interconnected. 
Specifically, the completion time depends on the size of composed model, which affects the determination of local update frequency.
Meanwhile, the local update frequency also affects the training adequacy of each parameter in coefficient.
Therefore, \textit{it is necessary yet challenging to jointly assign proper coefficient and local update frequency for each client.}
Our contributions are summarized as follows:
\begin{itemize}
    \item We propose a lightweight FL framework, called Heroes, which overcomes the challenges of resource limitation and client heterogeneity through enhanced neural composition and adaptive local update. Besides, a theoretical convergence analysis is provided for Heroes.
    \item  Guided by the convergence bound, we design a greedy-based algorithm to adaptively assign the proper coefficients and local update frequencies for participating clients based on both their heterogeneous capabilities and resource budgets.
    \item The performance of Heroes is evaluated through extensive experiments and the results demonstrate that Heroes can reduce the traffic consumption by about 72.05\% and provide up to 2.97$\times$ speedup for the training process compared to the baselines.
\end{itemize}

%% file: content/motivation.tex
\subsection{Federated Learning}
Considering a client set $\mathcal{N} = \{1, 2, \cdots, N \}$ coordinated by the PS, each client $n \in \mathcal{N}$ holds its local dataset $D_{n} = \{\zeta_{i}^{n}\}_{i=1}^{|D_{n}|}$, where $\zeta_{i}^{n}$ denotes a data sample from $D_{n}$.
Further, we represent the loss function as $\mathcal{L}(\mathbf{x}; \zeta)$, which measures how well the model $\mathbf{x}$ performs on data sample $\zeta$.
Therefore, the local loss function of client $n$ is defined as:
\begin{equation}
    F_{n}(\mathbf{x}) \triangleq \mathbb{E}_{\zeta \thicksim D_{n}} [\mathcal{L}(\mathbf{x}; \zeta)] 
\end{equation}

The global loss function is a linear combination of all $N$ clients, and the goal of FL is to train a high-quality model $\mathbf{x}^{*}$ with minimum global loss function, which is defined as:
\begin{equation} \label{eq:fl_goal}
    \mathbf{x}^{*} \triangleq  \mathop{\arg\min}_{\mathbf{x}} F(\mathbf{x}) = \mathop{\arg\min}_{\mathbf{x}} \frac{1}{N} \sum_{n=1}^{N} F_{n} (\mathbf{x})
\end{equation}

Suppose there are $H$ rounds of training in total. 
In each round $h \in \{1, 2, \cdots, H\}$, the PS randomly selects a set of clients $\mathcal{N}^{h} \subseteq  \mathcal{N}$ to participate in training and sends the fresh global model $\mathbf{x}^{h}$ to the specified clients, where $|\mathcal{N}^{h}| = K$.
Then, each client $n \in \mathcal{N}^{h}$ updates the global model $\mathbf{x}^{h}$ over its local dataset $D_{n}$ for $\tau$ times, where each update is regarded as one local iteration and $\tau$ represents the local update frequency.
Let $\mathbf{x}^{h}_{n} (t)$ denote the local model of client $n$ at iteration $t$ in round $h$.
For the mini-batch stochastic gradient descent (SGD) \cite{yu2019parallel} algorithm, a local iteration can be expressed as follows:
\begin{equation}
    \mathbf{x}^{h}_{n} (t+1) = \mathbf{x}^{h}_{n} (t) - \eta \nabla F_{n} (\mathbf{x}^{h}_{n} (t); \xi^{n})
\end{equation}
where $\xi^{n}$ denotes a random data batch from the local dataset $D_{n}$ and $\eta$ is the learning rate.
Finally, the PS collects the updated local models from the participating clients and aggregates them to the latest global model for further training, \ie, $ \mathbf{x}^{h+1} = \frac{1}{N} \sum_{n=1}^{N} \mathbf{x}^{h}_{n} (\tau)$.

\subsection{Enhanced Neural Composition}
To make full use of the limited resources, it is necessary to adjust the complexity of local model for each client based on its resource budgets \cite{diao2020heterofl, horvath2021fjord, mei2022resource}.
Herein, we follow \textit{Flanc} \cite{mei2022resource} to innovatively propose an enhanced neural composition technique to construct the models in different widths (\ie, the number of hidden channels in each weight) based on low-rank factorization \cite{phan2020stable}. Concretely, since the weights in DNNs are usually over-parameterized \cite{zou2019improved}, each layer's weight (\eg, convolution, fully connection) can be approximated as the product of two low-rank tensors, named neural basis $\mathbf{v}$ and coefficient $\mathbf{u}$.
For example, let $\mathbf{w} \in \mathbb{R}^{k^{2} \times I \times O}$ represent a convolution weight, with kernel size $k$, input channel number $I$ and output channel number $O$.
Let $p \in \{1, 2, \cdots, P\}$ denote the weight width, where the shape of $p$-width weight $\mathbf{w}_{p}$ is $k^{2} \times pI \times pO$.
Accordingly, $\mathbf{w}_{p}$ is approximated as follows:
\begin{equation}
    \mathbf{w}_{p} \approx \mathbf{v} \cdot \mathbf{u}_{p}, \qquad \mathbf{v} \in \mathbb{R}^{k^{2} \times I \times R}, \mathbf{u} \in \mathbb{R}^{R \times (p \times p O)}
\end{equation}

When $k=1$, the above format represents the approximation of fully connection layer's weight.
The weight width is controlled by adjusting the size of coefficient, while the size of neural basis is constant.
Specifically, the complete coefficient is divided into $P^{2}$ blocks, where the shape of each block is $R \times O$.
We select $p^{2}$ blocks from the complete coefficient to form the reduced coefficient and compose (\ie, multiply) it with neural basis into a $p$-width weight.

To ensure sufficient training of every parameter in the global model, we control different coefficient blocks to be trained evenly.
Thus, the selected blocks are currently the least trained ones.
Notably, we measure each block's training adequacy by the total number of local iterations it has experienced on all clients since round 1, \ie, total update times.
In general, we balance the total update times of different coefficient blocks to ensure that each block is fully trained.

\begin{figure}[t]
    \centering
    \includegraphics[width=0.45\textwidth]{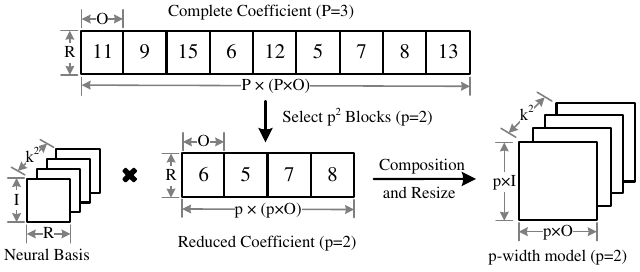}
        \caption{The demonstration of constructing a $p$-width weight by the enhanced neural composition technique ($P$=3, $p$=2).} \label{fig:neural_composition}\vspace{-0.5cm}
\end{figure}

For example, as illustrated in Fig. \ref{fig:neural_composition}, the coefficient $\mathbf{u} \in \mathbb{R}^{R \times (9 \times O)}$ is divided into 9 blocks (\ie, $P$=3) and the number in each block represents its current total update times.
To obtain a 2-width weight, we first extract the least trained 4 blocks (with the total update times of 6, 5, 7 and 8, respectively) from the complete coefficient and combine them into the reduced coefficient $\hat{\mathbf{u}} \in \mathbb{R}^{R \times (4 \times O)}$.
Then, $\hat{\mathbf{u}}$ is composed with the neural basis $\mathbf{v} \in \mathbb{R}^{k^2 \times I \times R}$ into an intermediate tensor whose shape is $k^2 \times I \times (4 \times O)$.
Finally, a 2-width weight $\hat{\mathbf{x}} \in \mathbb{R}^{k^2 \times (2 \times I) \times (2 \times O)}$ is obtained by reshaping this intermediate tensor.

\begin{table}[h]
\centering
\setlength{\abovecaptionskip}{10pt}%
\setlength{\belowcaptionskip}{0pt}%
\footnotesize
\caption{Training performance within given resource constraints.}
\label{table:motivation_training_performance_completion_time}
\begin{tabular}{|c|c|c|c|c|}
\hline
\multirow{2}{*}{FL Schemes} & \multicolumn{2}{c|}{Traffic} &  \multicolumn{2}{c|}{Time} \\
\cline{2-5}
& 30GB & 60GB & 20,000s & 40,000s \\
\hline
MP \cite{diao2020heterofl} & 34.89\% & 48.92\% & 39.72\% & 50.75\% \\
\hline
Original NC \cite{mei2022resource} & 42.22\% & 51.89\% & 43.25\% & 49.37\% \\
\hline
\textbf{Enhanced NC} & \textbf{59.76\%} & \textbf{64.42\%} & \textbf{58.69\%} & \textbf{62.81\%}  \\
\hline
\end{tabular}
\end{table}

Compared to the FL schemes based on model pruning \cite{diao2020heterofl}, the neural composition technique enables every parameter in the global model to benefit from the full range of knowledge through the shared neural basis, improving the training performance.
Different from \textit{Flanc} \cite{mei2022resource} with original neural composition, enhanced neural composition allows the coefficient of all sizes to get fully trained, accelerating the training process.
To verify this improvement, we simulate a FL system with 100 clients to train the standard ResNet-18 model \cite{he2016deep} over the ImageNet dataset \cite{russakovsky2015imagenet}.
The results in Table \ref{table:motivation_training_performance_completion_time} indicate that with the given traffic consumption or completion time, the enhanced neural composition can improve the global model's test accuracy by about 16.29\% on average compared with the FL schemes based on model pruning (MP) \cite{diao2020heterofl} and original neural composition (NC) \cite{mei2022resource}.

\subsection{Adaptive Local Update}
\begin{figure}[t]
	\centering
	\subfigure[Fixed and Identical Frequency]{
		\includegraphics[width=1.6in]{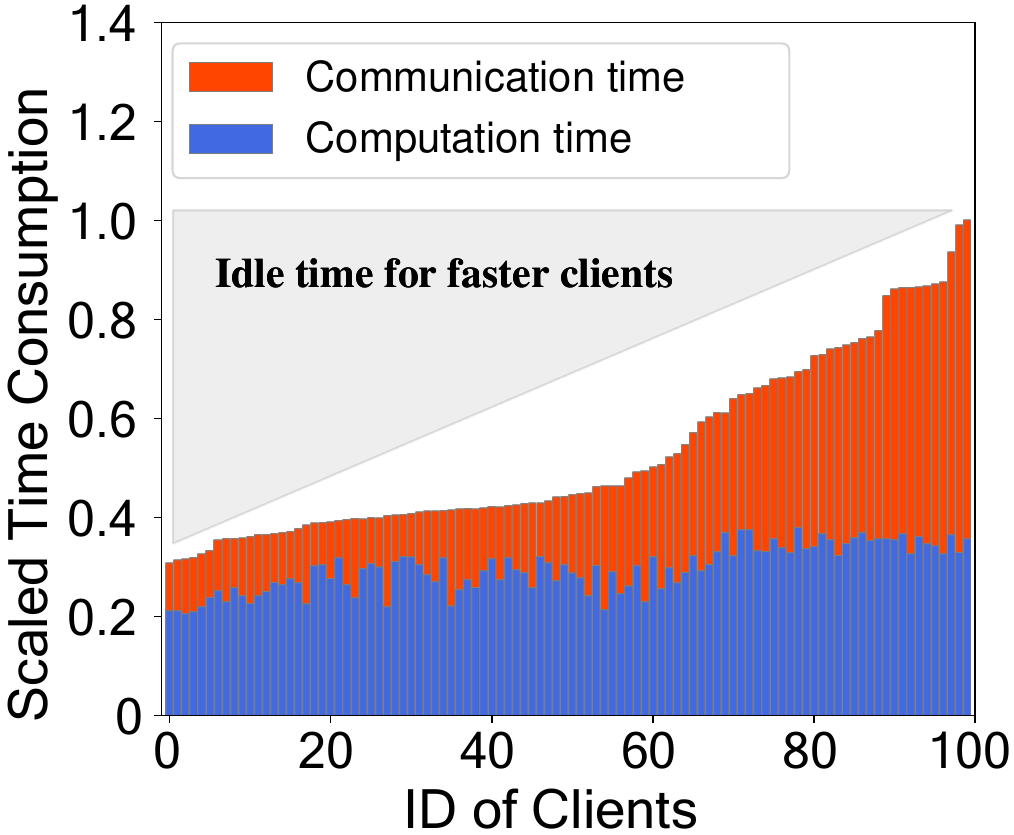}\label{fig:round_time_noadloc}
	}
	\subfigure[Adaptive Frequency]{
		\includegraphics[width=1.6in]{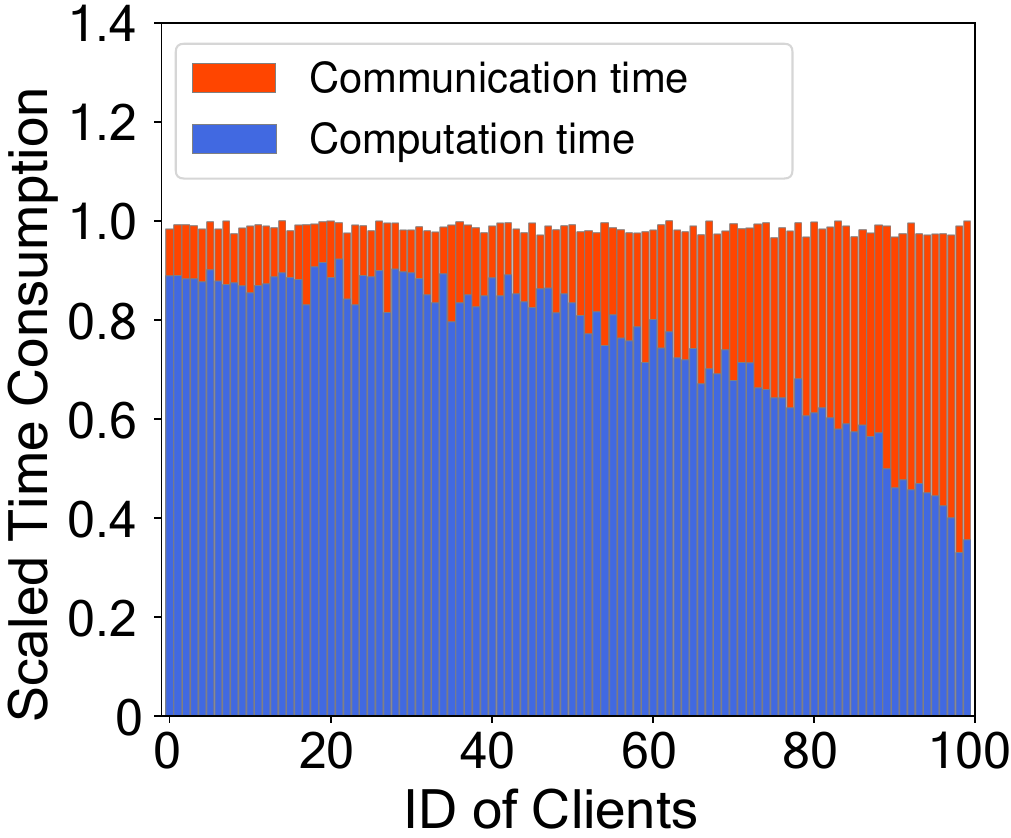}\label{fig:round_time_adloc}
	}
 \vspace{-2mm}
	\caption{Ranked clients' completion time in one training round.}\label{fig:round_time}
	\vspace{-3mm}
\end{figure}

In most synchronous FL schemes \cite{ li2021talk, mei2022resource}, the clients' local update frequencies are identical and fixed in each round.
However, due to client heterogeneity, strong clients have to wait for weak ones (\ie, stragglers) for global aggregation, incurring non-negligible waiting time and reducing the training efficiency significantly \cite{liu2023adaptive, wang2020towards}. 
To evaluate the negative impacts of stragglers, we record the completion time for one training round of each client in the simulated FL system.
As shown in Fig. \ref{fig:round_time_noadloc}, the strongest client completes one training round four times faster than the weakest client.
In other words, about 70\% of the strongest client's time is idle and wasted.

To reduce clients' idle waiting, some research \cite{li2020federated, xu2022adaptive, li2022pyramidfl} proposes adjusting the local update frequencies for different clients to balance their completion time.
On the one hand, the weaker clients perform fewer local iterations to reduce the impact of stragglers.
On the other hand, the stronger clients utilize the idle time to perform more local iterations, which helps the model converge faster.
We illustrate the completion time of each client with proper local update frequency in Fig. \ref{fig:round_time_adloc}.
It can be observed that almost every client's time is fully utilized without idle waiting.

However, the neural composition and the local update are interconnected, and it is difficult to jointly assign proper coefficient blocks and local update frequencies for different clients.
Specifically, both the computation time and the communication time of a specific client depend on the number of coefficient blocks, which affects the determination of its local update frequency.
Meanwhile, since each client trains different blocks, various local update frequencies will also affect the balance of the blocks' total update times.

%% file: content/framework.tex
To enhance the training efficiency and resource utilization for FL, we propose a lightweight FL framework, called Heroes, which integrates the benefits of enhanced neural composition and adaptive local update.
Specifically, in Heroes, there are three main phases in each round as follows.

\subsubsection{\textbf{Tensors and Frequency Assignment}}
The PS first determines the model width $p_{n}^{h}$ for each client $n$ in round $h$.
As studied in \cite{diao2020heterofl, mei2022resource}, the smaller the model width, the less the required computation resource.
To alleviate the performance degradation, Heroes increases each client's model width as much as possible within its resource budget.
Then, for each client $n$, Heroes selects $(p_{n}^{h})^{2}$ coefficient blocks with the least total update times as the reduced coefficient $\hat{\mathbf{u}}_{n}^{h}$.
Finally, Heroes assigns a proper local update frequency $\tau_{n}^{h}$ to each client $n$ by the algorithm proposed in Section \ref{subsec:algorithm_description} to minimize the idle waiting time and balance the training across different coefficient blocks. 

\subsubsection{\textbf{Local Training}}
In round $h$, each client $n \in \mathcal{N}^{h}$ first downloads the latest neural basis $\mathbf{v}^{h}$ and the reduced coefficient $\hat{\mathbf{u}}_{n}^{h}$ from the PS, then composes them into the local model $\hat{\mathbf{x}}_{n}^{t}$.
Each local model $\hat{\mathbf{x}}_{n}^{t}$ is trained over the local dataset $D_{n}$ for $\tau_{n}^{h}$ iterations.
Let $\bar{\mathbf{x}}_{n}^{h}$ represent the updated local model of client $n$ in round $h$.
After the local training, $\bar{\mathbf{x}}_{n}^{h}$ is be decomposed into the updated neural basis $\bar{\mathbf{v}}_{n}^{h}$ and coefficient $\bar{\mathbf{u}}_{n}^{h}$, \ie, $\bar{\mathbf{x}}_{n}^{h} \approx \bar{\mathbf{v}}_{n}^{h} \cdot \bar{\mathbf{u}}_{n}^{h}$.
Since the size of low-rank tensors is inherently smaller than that of original model, each client $n$ uploads $\bar{\mathbf{v}}_{n}^{h}$ and $\bar{\mathbf{u}}_{n}^{h}$, instead of $\bar{\mathbf{x}}_{n}^{h}$, to the PS for global aggregation, which further saves the limited bandwidth.

\subsubsection{\textbf{Global Aggregation}}
\begin{figure}[t]
    \centering
    \includegraphics[width=0.45\textwidth]{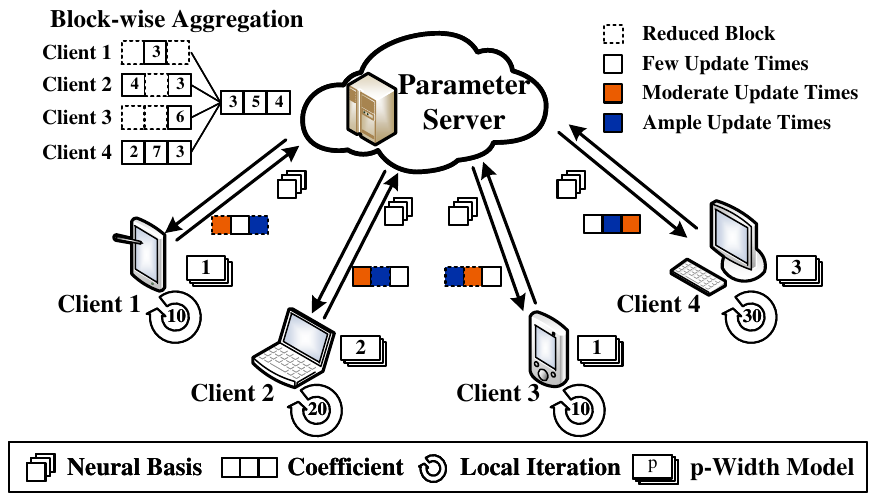}
    \caption{The demonstration of the proposed framework.} \label{fig:framework}\vspace{-5mm}
\end{figure}

In round $h$, upon receiving the updated neural basis and coefficient from all the participating clients in $\mathcal{N}^{h}$, the PS performs global aggregation to obtain the latest basis $\mathbf{v}^{h+1}$ and coefficient $\mathbf{u}^{h+1}$ for the next round of training.
For neural basis, Heroes directly averages the updated ones from all participating clients, \ie, $\mathbf{v}^{h+1} = \frac{1}{K} \sum_{n=1}^{K} \bar{\mathbf{v}}_{n}^{h}$.
For coefficient, Heroes performs the block-wise aggregation.
Specifically, let $i \in \{1, 2, \cdots, P^{2} \}$ represent the block index and $\mathcal{N}_{i}^{h}$ denote the set of clients that train the $i$-th coefficient block in round $h$. 
The latest $i$-th coefficient block $\mathbf{u}^{h+1, i}$ is obtained as follows:
\begin{equation}
    \mathbf{u}^{h+1, i} = \frac{1}{|\mathcal{N}_{i}^{h}|} \sum_{n \in \mathcal{N}_{i}^{h}} \bar{\mathbf{u}}_{n}^{h, i}
\end{equation}
where $\bar{\mathbf{u}}_{n}^{h, i}$ represents the $i$-th coefficient block updated by client $n$ in round $h$.
For instance, as shown in Fig. \ref{fig:framework}, the leftmost block is trained by two clients (\ie, 2 and 4), thus its value is $3 = \frac{1}{2} \times (4+2)$.

For better explanation of Heroes, we give a demonstration of Heroes in Fig. \ref{fig:framework}. 
Four heterogeneous clients participate in training, which are divided into three levels by their computation power, \ie, \textit{weak} smartphones (clients 1 and 3), \textit{medium} laptop (client 2) and \textit{powerful} PC (client 4).
The complete coefficient contains three blocks.
Heroes selects one coefficient block for the \textit{weak} clients 1 and 3, and two blocks for \textit{medium} client 2.
The \textit{powerful} client 4 utilizes all three blocks for model training.
We adopt the block's color to denote the amount of training it has obtained.
Specifically, the total update times of blue blocks is ample, while that of orange blocks and white blocks is moderate and few, respectively.
To ensure sufficient training for every block, the selected blocks are the least trained ones currently.
Besides, Heroes assigns different local update frequencies for these four clients to balance their completion time.
For example, the \textit{weak} client 1 trains the model much slower than the \textit{powerful} client 4.
Thus, client 1 performs fewer local iterations (\eg, 10) to alleviate the straggler effect, while client 4 performs more local iterations (\eg, 30) to make more contributions to the global model.
In addition to the balance of clients' completion time, the balance of coefficient blocks' total update times is also considered for the determination of each client's local update frequency.


%% file: content/analysis.tex
In this section, we provide the convergence analysis of the proposed framework.
We first state the following assumptions, which are standard in non-convex optimization problems and widely used in the analysis of previous works \cite{jiang2023computation}.

\noindent\textbf{Assumption 1.} \textbf{(\textit{Smoothness})} The local objective function $F_{n}$ of each client $n$ is smooth with modulus $L$:
\begin{equation}
    \! \!F_{n}(\mathbf{y})\! - \!F_{n}(\mathbf{x}) \leq \frac{L}{2} \| \mathbf{y} - \mathbf{x} \|^{2} + \langle \nabla F_n(\mathbf{y}), \mathbf{y} - \mathbf{x} \rangle \quad \forall n, \mathbf{x}, \mathbf{y} 
\end{equation}

\noindent\textbf{Assumption 2.} \textbf{(\textit{Bounded Variance})} 
Let $\xi_{n}$ denote a random data batch sampled from client $n$'s local dataset $D_{n}$. 
There exists a constant $\sigma > 0$ such that the variance of stochastic gradients at each client is bounded by:
\begin{equation}
    \mathbb{E} [ \| \nabla F_{n}(\mathbf{x}; \xi_{n}) - \nabla F_{n} (\mathbf{x}) \|^{2} ] \leq \sigma^{2} \quad \forall \mathbf{x}, n, \xi_{n} 
\end{equation}
\vspace{0.1mm}
\noindent\textbf{Assumption 3.} \textbf{(\textit{Bounded Gradients})} There exists a constant $G>0$, such that the stochastic gradients at each client $n$ are bounded by:
\begin{equation}
    \mathbb{E} [ \| \nabla F_{n}(\mathbf{x}; \xi_{n}) \|^{2} ] \leq G^{2} \quad \forall \mathbf{x}, n, \xi_{n}  
\end{equation}

In order to analyze the convergence bound of the global loss function after $H$ rounds, we present the following three important lemmas.

\noindent\textbf{Lemma 1.} According to Assumption 3, the deviations between global model $\mathbf{x}^{h}$ and local model $\hat{\mathbf{x}}^{h}_{n}(t-1)$ is bounded as:
\begin{align}
    & \! \! \! \! \sum_{t=1}^{\tau} \sum_{n=1}^{N} \mathbb{E} [\| \mathbf{x}^{h}\! -\! \hat{\mathbf{x}}_{n}^{h}(t\! -\! 1) \|^{2} ] \!\leq\! \frac{2 \tau^{3} \eta^{2} G^{2} N }{3} \! + \! 2 \tau \sum_{n=1}^{N} \alpha_{n}^{h} 
\end{align}
where $\alpha_{n}^{h}$ represents the model error induced by reducing coefficient for client $n$ in round $h$ (\ie, $ \alpha_{n}^{h} \triangleq  \|\mathbf{u}^{h} - \hat{\mathbf{u}}^{h}_{n} \|^{2}$) and $\tau = \mathop{\max_{n, h}} \{ \tau_{n}^{h} \}$.
 
\noindent\textbf{Lemma 2.} Combining Assumptions 1-2 and Lemma 1, the difference between global models in two successive rounds can be bounded as:
\begin{align}
    &  \mathbb{E} [\| \mathbf{x}^{h+1} - \mathbf{x}^{h} \|^{2}] \leq  3 \eta^{2} \tau^{2} \mathbb{E} [\|  \nabla F_{n} (\mathbf{x}^{h})  \|^{2}]  \notag \\
    & +  \frac{6 \eta^{2} \tau^{2} L^{2}}{N} \sum_{n=1}^{N} \alpha_{n}^{h} + 2 \tau^{4} \eta^{4} G^{2} L^{2} + 3 \eta^{2} \tau^{2} \sigma^{2} \label{eq:lemma_2}
\end{align}

\noindent\textbf{Lemma 3.} Under Assumption 1 and Lemma 1, the proposed framework ensures that:
\begin{align}
    &\mathbb{E}[\langle \nabla F(\mathbf{x}^{h}), \mathbf{x}^{h+1} - \mathbf{x}^{h} \rangle ]  \notag \\
    \leq & - \frac{\tau \eta}{2} \mathbb{E}[\| \nabla F(\mathbf{x}^{h}) \|^{2}] + \frac{L^{2} \tau \eta}{N} \sum_{n=1}^{N} \alpha_{n}^{h} + \frac{\tau^{3} \eta^{3} L^{2} G^{2}}{3} \label{eq:lemma_3}
\end{align}

Let these assumptions and lemmas hold, then the convergence bound can be obtained as follows.

\noindent\textbf{Theorem 1.} If the learning rate satisfies $\eta \leq \frac{1}{6L\tau}$, the mean square gradient after $H$ rounds can be bounded as follows:

\begin{align}
    & \frac{1}{H} \sum_{h=0}^{H-1} \mathbb{E}[\| \nabla F(\mathbf{x}^{h}) \|^{2}] \leq \frac{4}{H \eta \tau} (F(\mathbf{x}^{0}) - F(\mathbf{x}^{*})) \notag \\
    &  + \frac{6 L^{2}}{H N} \sum_{h=0}^{H-1} \sum_{n=1}^{N} \alpha_{n}^{h} + \frac{L \eta \tau }{3} ( G^{2} + 18 \sigma^{2} ) \label{eq:theorem_1}
\end{align}
where $\mathbf{x}^{*}$ denotes the optimal global model. 

\noindent\textit{Proof. } With the smoothness assumption, we have:
\begin{align}
    & \mathbb{E}[F(\mathbf{x}^{h+1})] - \mathbb{E}[F(\mathbf{x}^{h})] \notag \\
    \leq & \frac{L}{2} \mathbb{E} [\| \mathbf{x}^{h+1} - \mathbf{x}^{h} \|^{2}] + \mathbb{E}[\langle \nabla F(\mathbf{x}^{h}), \mathbf{x}^{h+1} - \mathbf{x}^{h} \rangle ] \label{eq:theorem_1_smooth} \\ 
    \leq & \frac{\tau \eta}{2}(3 \tau \eta L - 1) \mathbb{E}[\| \nabla F(\mathbf{x}^{h}) \|^{2}] + \frac{L^{2}\tau\eta}{N}(3 L \tau \eta + 1) \sum_{n=1}^{N} \alpha_{n}^{h} \notag \\
    & + \tau^{4} \eta^{4} L^{3} G^{2} + \frac{3 \eta^{2} \tau^{2} \sigma^{2} L}{2} + \frac{\tau^{3} \eta^{3} L^{2} G^{2}}{3} \label{eq:theorem_1_taking_lemmas_2_3}
\end{align}
where Eq. \eqref{eq:theorem_1_taking_lemmas_2_3} is obtained by inserting Eq. \eqref{eq:lemma_2} and Eq. \eqref{eq:lemma_3} into Eq. \eqref{eq:theorem_1_smooth}. Summing over the round $h \in \{0, 1, \cdots, H-1 \}$ on the both sides of Eq. \eqref{eq:theorem_1_taking_lemmas_2_3}, we have:
\begin{flalign}
    & \mathbb{E}[F(\mathbf{x}^{H})] - \mathbb{E}[F(\mathbf{x}^{0})] \leq \tau^{4} \eta^{4} L^{3} G^{2} H  \notag \\
    & + \frac{\tau \eta}{2}(3 \tau \eta L - 1) \sum_{h=0}^{H-1} \mathbb{E}[\| \nabla F(\mathbf{x}^{h}) \|^{2}] + \frac{3 \eta^{2} \tau^{2} \sigma^{2} H L}{2} \notag \\
    & + \frac{L^{2}\tau\eta}{N}(3 L \tau \eta + 1) \sum_{h=0}^{H-1} \sum_{n=1}^{N} \alpha_{n}^{h} + \frac{\tau^{3} \eta^{3} L^{2} G^{2} H}{3} \label{eq:theorem_1_sum}
\end{flalign}
When the learning rate satisfies $\eta \leq \frac{1}{6L\tau}$, we can shift the terms in Eq. \eqref{eq:theorem_1_sum} as follows:
\begin{align}
    & \frac{\tau \eta}{4} \sum_{k=0}^{H-1} \mathbb{E}[\| \nabla F(\mathbf{x}^{h}) \|^{2}] \leq \mathbb{E}[F(\mathbf{x}^{0})] - \mathbb{E}[F(\mathbf{x}^{H})] \notag \\
    & + \frac{3 L^{2} \tau \eta}{2 N} \sum_{h=0}^{H-1} \sum_{n=1}^{N} \alpha_{n}^{h} + \frac{\tau^{2} \eta^{2} H L}{12} ( G^{2} + 18 \sigma^{2} ) \label{eq:theorem_1_shift}
\end{align}

Finally, we divide both sides of Eq. \eqref{eq:theorem_1_shift} by $\frac{H \eta \tau}{4}$ to derive the convergence bound in Eq. \eqref{eq:theorem_1} of Theorem 1.

Thus, we complete the convergence analysis of the proposed framework under the non-convex setting.
The convergence bound is proportional to the coefficient reducing error. 
The larger the coefficient, the smaller the reducing error, leading to a tighter convergence bound.
Besides, the bound also depends on the local update frequency $\tau$, indicating that we can obtain better convergence performance by determining the local update frequency $\tau$ properly.

%% file: content/algorithm.tex

\subsection{Problem Formulation}

In this section, we define the joint optimization problem of neural composition and local update frequency in FL training.
Let $G(\mathbf{v} \cdot \mathbf{u})$ represent the number of floating-point operations (FLOPs) required to perform one local iteration for the composed model, and $G(\mathbf{v} \cdot \mathbf{u})$ depends on the size of coefficient $\mathbf{u}$.
The more blocks the coefficient $\mathbf{u}$ includes, the wider the composed model $\mathbf{x} = \mathbf{v} \cdot \mathbf{u}$ and the more FLOPs required for model training.  
We formulate the time cost for one local iteration of client $n$ in round $h$ as follows:

\begin{equation}
    \mu_{n}^{h} = G(\mathbf{v}^{h}_{n} \cdot \hat{\mathbf{u}}^{h}_{n}) / q_{n}^{h}
\end{equation}
where $q_{n}^{h}$ represents the speed to process the floating-operations of client $n$ in round $h$.

Since the download bandwidth is usually much faster than the upload bandwidth in typical WANs \cite{zhan2020incentive}, the download time of tensors can be negligible and we mainly focus on the upload time. 
Let $b_{n}^{h}$ denote the upload bandwidth of client $n$ in round $h$ and $E$ measure the size of a tensor.
We formulate the communication time of client $n$ in round $h$ as:
\begin{equation}
    \nu_{n}^{h} = [E(\bar{\mathbf{v}}_{n}^{h}) + E(\bar{\mathbf{u}}_{n}^{h})] / b_{n}^{h}
\end{equation}

Let $T_{n}^{h}$ denote the completion time of client $n$ in round $h$, which consists of the time cost for $\tau_{n}^{h}$ local iterations and the communication time.
Due to the synchronization barrier of FL, the completion time of round $h$ depends on the slowest client, which can be defined as:
\begin{equation}
    T^{h} = \mathop{\max}_{n \in \mathcal{N}^{h}}  T_{n}^{h} = \mathop{\max}_{n \in \mathcal{N}^{h}} (\tau_{n}^{h} \cdot \mu_{n}^{h} + \nu_{n}^{h}) 
\end{equation}

We adopt the average waiting time for participating clients in $\mathcal{N}^{h}$ to measure the impact of synchronization barrier in round $h$, which is defined as follows:
\begin{equation}
    \mathcal{W}^{h} = \frac{1}{K} \sum_{n \in \mathcal{N}^{h}} (T^{h} - T_{n}^{h})
\end{equation}

Let $c_{i}^{h}$ denote the total update times of $i$-th coefficient block, where $i \in \{1, 2, \cdots, P^{2} \}$.
The training consistency between different coefficient blocks in round $h$ can be reflected by the variance of set $\{ c_{i}^{h} | \forall i\}$, denoted as:
\begin{equation}
    \mathcal{V}^{h} = \frac{1}{P^{2}} \sum_{i=1}^{P^{2}} ( c_{i}^{h} - \frac{1}{P^{2}} \sum_{j=1}^{P^{2}} c_{j}^{h})^{2}
\end{equation}


In each round $h$, we aim to select appropriate $(p_{n}^{h})^{2}$ coefficient blocks and determine the optimal local update frequency $\tau_{n}^{h}$ for each participating client $n \in \mathcal{N}^{h}$, so as to accelerate the training process of FL.
Accordingly, we define the optimization problem as follows:

\centerline{$\mathop{\min} \sum_{h=1}^{H} T^{h}$}
\begin{equation} \label{eq:problem_formulation}
    s.t.
    \begin{cases} 
    F(\mathbf{v}^{H} \cdot \mathbf{u}^{H}) \leq \epsilon  \\
    \mathcal{W}^{h} \leq \rho, & \forall h \\
    \mathcal{V}^{h} \leq \delta, & \forall h \\
    p_{n}^{h} \in \{ 1, 2, \cdots, P \}, & \forall n, h
    \end{cases} 
\end{equation}

The first inequality expresses the convergence requirement, where $\epsilon$ is the convergence threshold of the training loss after $H$ rounds. 
The second set of inequalities indicates that the average waiting time of participating clients in each round $h$ should not exceed the given threshold $\rho$.
The third set of inequalities bounds the variance of all coefficient blocks' total update times in each round $h$, ensuring the balanced training among blocks. 
The fourth set of inequalities tells the feasible range of the number of coefficient blocks.
The object of this optimization problem is to minimize the completion time of the FL training with the performance requirements (\eg, model convergence, average waiting time).

\begin{algorithm}[t]
	\caption{ Procedure at the PS} \label{alg:ps}
    \algnotext{EndFor}
    \algnotext{EndWhile}
    \algnotext{EndIf}
    \algnotext{label}
	\algorithmicrequire{ completion time budget $T^{max}$, maximum time cost for one local iteration $\mu^{max}$, maximum model width $P$, waiting time bound $\rho$.}\\
	\algorithmicensure{ convergenced neural basis $\mathbf{v}^{H}$ and coefficient $\mathbf{u}^{H}$.}
	\begin{algorithmic}[1]
		\State Initialize $\mathbf{v}^{0}$ and $\mathbf{u}^{0}$ as random tensors;
        \State Initialize $h \leftarrow 0$, $T \leftarrow 0$, $c_{i}^{h} \leftarrow 0$ ($\forall i \in [P^{2}]) $;
		\While { $T \leq T^{max}$ }
        \State Collect the status information of network and clients;
        \State Randomly sample $K$ participating clients, \ie, $\mathcal{N}^{h}$;
		\For {each client $n \in  \mathcal{N}^{h}$}
  \label{alg:ps_determine_block_number_begin}
        \State Set $\mu_{n}^{h} \leftarrow 0$ and $p_{n}^{h} \leftarrow 1$; 
        \While{$\mu_{n}^{h} < \mu^{max}$ and $p_{n}^{h} < P$}
        \State Estimate the local iteration time $\mu_{n}^{h}$;
        \State Set $p_{n}^{h} \leftarrow p_{n}^{h} + 1$;
        \EndWhile
        \State Estimate the communication time $\nu_{n}^{h}$; \label{alg:ps_determine_block_number_end}
		\EndFor
        \For { each client $n \in  \mathcal{N}^{h}$} \label{alg:ps_select_l_begin}
        \State Solve Eq. \eqref{eq:approximated_client_n_completion_time} to obtain $T_{n}$, $\tau_{n}$ and $T_{n}^{h}$; 
        \EndFor 
        \State Select the fastest client $l \leftarrow \mathop{\arg\min}_{n \in \mathcal{N}^{h}} T_{n}$; \label{alg:ps_select_l_end}
        \State Set $\tau_{l}^{h} \leftarrow \tau_{l}$ and $T \leftarrow T + T_{l}^{h}$;\label{alg:ps_determine_other_tau_begin}
        \For { each client $n \in  \mathcal{N}^{h}$} 
        \If {$n \neq l$}
        \State Obtain the interval $[\tau_{a}, \tau_{b}]$ by Eq. \eqref{eq:approximate_waiting_time};  
        \State Search $\tau_{n}^{h} \in [\tau_{a}, \tau_{b}]$ to minimize $\mathcal{V}^{h}$; \label{alg:ps_determine_other_tau_end}
        \EndIf
        \State Select the least trained $(p_{n}^{h})^{2}$ blocks to form $\hat{\mathbf{u}}^{h}_{n}$;
        \For {each coefficient block $i$ in $\hat{\mathbf{u}}_{n}^{h}$ }
        \State Update $c_{i}^{h} \leftarrow c_{i}^{h} + \tau_{n}^{h}$;
        \EndFor
        \State Send $\mathbf{v}^{h}$, $\hat{\mathbf{u}}_{n}^{h}$, $\tau^{h}_{n}$ to client $n$;
		\State Receive $L_{n}$, $\sigma_{n}^{2}$, $G_{n}^{2}$, $\bar{\mathbf{v}}^{h}_{n}$, $\bar{\mathbf{u}}_{n}^{h}$ from client $v_{n}$;
        \EndFor 
        \State Aggregate estimated variables to $L$, $\sigma^{2}$ and $G^{2}$; 
        \label{alg:ps_global_aggregation_begin}
        \
        \State Aggregate basis and coefficient to $\mathbf{v}^{h+1}$ and $\mathbf{u}^{h+1}$;\label{alg:ps_global_aggregation_end}
        \State Set $h \leftarrow h + 1$ and $c_{i}^{h} \leftarrow c_{i}^{h-1}$ ($\forall i \in [P^{2}]$);
		\EndWhile 
        \State Set $H \leftarrow h-1$ ,then send $\mathbf{v}^{H}$ and $\mathbf{u}^{H}$ to all clients.
	\end{algorithmic}
\end{algorithm}

\subsection{Preliminaries for Algorithm Design}
To solve the optimization problem in Eq. \eqref{eq:problem_formulation}, we first approximate the convergence bound. 
Specifically, we adopt an upper bound $\beta^{2}$ for the coefficient reducing error, where $\alpha_{n}^{h} \leq \beta^{2}$. 
Besides, the minimum loss value is approximated as zero, \ie, $F(\mathbf{x}^{*}) = 0$. 
Accordingly, the convergence bound in Eq. \eqref{eq:theorem_1} is formulated as follows:
\begin{equation}
    G(H, \tau) = \frac{4}{H \eta \tau} F(\mathbf{x}^{0}) + \frac{L \eta \tau }{3} ( G^{2} + 18 \sigma^{2}) + 6 L^{2} \beta^{2}
\end{equation}

Since $\tau$ is a positive variable, the convergence bound $G(H, \tau)$ is a convex function with respect to the local update frequency $\tau$.
It can be derived that $G(H, \tau)$ will decrease as $\tau$ increases when $\tau \leq \small{\sqrt{\frac{12 F(\mathbf{x}^{0})}{\eta^{2} H L (G^{2} + 18 \sigma^{2})}}} $.
On the contrary, the trend of $G(H, \tau)$ and $\tau$ is the opposite.

To balance the completion time of heterogeneous clients, we first select the fastest client $l$ in round $h$ and let the completion time of other clients be approximately equal to that of client $l$, which can be formulated as follows:
\begin{equation} \label{eq:approximate_waiting_time}
    0 \leq T_{l}^{h} - (\tau_{n}^{h} \cdot \mu_{n}^{h} + \nu_{n}^{h}) \leq \rho , \quad \forall n \in \mathcal{N}^{h}, \forall h
\end{equation} 

Notably, the completion time $T_{l}^{h}$ of client $l$ is the largest among that of all participating clients in round $h$.
Therefore, the total completion time can be denoted as follows:
\begin{equation}
    T(H, \tau) = \sum_{h=1}^{H} T_{l}^{h} = \sum_{h=1}^{H}(\tau_{l}^{h} \cdot \mu_{l}^{h} + \nu_{l}^{h}) 
\end{equation}

In order to minimize the convergence bound, 
we set the local update frequency $\tau_{l}^{h}$ of the fastest client $l$ in round $h$ as $\small{\sqrt{\frac{12 F(\mathbf{x}^{h})}{\eta^{2} H L (G^{2} + 18 \sigma^{2})}}}$. Thus, the optimization problem in Eq. \eqref{eq:problem_formulation} can be approximated as a univariate problem of finding the number of rounds $H$.

\centerline{$\mathop{\min} \sum_{h=1}^{H}(\tau_{l}^{h} \cdot \mu_{l}^{h} + \nu_{l}^{h})$ }
\begin{equation} \label{eq:approximate_problem_formulation}
    s.t.
    \begin{cases} 
    \tau_{l}^{h} = \sqrt{\frac{12 F(\mathbf{x}^{h})}{\eta^{2} H L (G^{2} + 18 \sigma^{2})}}, & \forall h  \\
    0 \leq T_{l}^{h} - (\tau_{n}^{h} \cdot \mu_{n}^{h} + \nu_{n}^{h}) \leq \rho , & \forall n \in \mathcal{N}^{h}, \forall h \\
    \mathcal{V}^{h} \leq \delta, & \forall h
    \end{cases} 
\end{equation}


\begin{algorithm}[t]
	\caption{Procedure at client $n$ in round $h$}\label{alg:client}
    \algnotext{EndFor}
    \algnotext{EndWhile}
    \algorithmicrequire{ global neural basis $\mathbf{v}^{h}$, reduced coefficient $\hat{\mathbf{u}}_{n}^{h}$, local update frequency $\tau^{h}_{n}$.} \\
	\algorithmicensure{ estimated variables $L_{n}$, $\sigma_{n}^{2}$ and $G_{n}^{2}$, updated neural basis $\bar{\mathbf{v}}^{h}_{n}$ and coefficient $\bar{\mathbf{u}}_{n}^{h}$}.
	\begin{algorithmic}[1]
        \State Receive $\mathbf{v}^{h}$, $\hat{\mathbf{u}}_{n}^{h}$ and $\tau^{h}_{n}$ from PS;
        \State Compose basis and coefficient $\hat{\mathbf{x}}^{h}_{n} \leftarrow \mathbf{v}^{h} \cdot \hat{\mathbf{u}}^{h}_{n}$;
        \State Initialize learning rate $\eta$ and set $\hat{\mathbf{x}}^{h}_{n}(0) \leftarrow \hat{\mathbf{x}}^{h}_{n}$;
		\For{each local iteration $t \in \{1, 2,...,\tau_{n}^{h}\}$}
		\State Update $\hat{\mathbf{x}}_{n}^{h} (t) \leftarrow \hat{\mathbf{x}}_{n}^{h} (t-1)-\eta \nabla F_{n} (\hat{\mathbf{x}}_{n}^{h} (t-1); \xi_{n}^{h})$;
		\EndFor
        \State Set $\bar{\mathbf{x}}_{n}^{h} \leftarrow \hat{\mathbf{x}}_{n}^{h} (\tau^{h}_{n})$;
        \State Estimate $L_{n} \leftarrow \|\nabla F_{n}(\bar{\mathbf{x}}_{n}^{h}) - \nabla F_{n}(\hat{\mathbf{x}}_{n}^{h}) \| / \|\bar{\mathbf{x}}_{n}^{h} - \hat{\mathbf{x}}_{n}^{h} \|$; \label{alg:client_estimate_begin}
		\State Estimate $\sigma_{n}^{2} \leftarrow\mathbb{E} [\|\nabla F_{n}(\hat{\mathbf{x}}_{n}^{h}, \xi^{n})-\nabla F_{n}(\hat{\mathbf{x}}_{n}^{h})\|^{2} ]$;
        \State Estimate $G_{n}^{2} \leftarrow\mathbb{E} [\|\nabla F_{n}(\hat{\mathbf{x}}_{n}^{h}, \xi^{n}) \|^{2} ]$; \label{alg:client_estimate_end}
		\State Decompose the updated local model $\bar{\mathbf{x}}^{h}_{n} \rightarrow \bar{\mathbf{v}}^{h} \cdot \bar{\mathbf{u}}^{h}_{n}$;
		\State Send $L_{n}$, $\sigma_{n}^{2}$, $G_{n}^{2}$, $\bar{\mathbf{v}}^{h}_{n}$ and $\bar{\mathbf{u}}_{n}^{h}$ to the PS.
	\end{algorithmic}
\end{algorithm}

\subsection{Algorithm Description} \label{subsec:algorithm_description}
In terms of Eq. \eqref{eq:approximate_problem_formulation}, we design a greedy-based control algorithm to adaptively assign proper coefficients and local update frequencies for heterogeneous clients. 
The proposed algorithm consists of both the PS and client sides, which are formally described in Alg. \ref{alg:ps} and Alg. \ref{alg:client}, respectively.
We will introduce the algorithm in detail by the order of workflow in a training round.

Firstly, at the beginning of each round $h$, the algorithm determines the model width $p_{n}^{h}$ for each client $n \in \mathcal{N}^{h}$ (Lines \ref{alg:ps_determine_block_number_begin}-\ref{alg:ps_determine_block_number_end} of Alg. \ref{alg:ps}).
To minimize the reducing error, the algorithm greedily adds the coefficient blocks for each client as many as possible within a given maximum iteration time $\mu^{max}$ or the model width $p_{n}^{h}$ reaches the maximum value $P$.

Secondly, the algorithm selects the fastest client with the least completion time (Lines \ref{alg:ps_select_l_begin}-\ref{alg:ps_select_l_end} of Alg. \ref{alg:ps}).
Specifically, for each participating client $n \in \mathcal{N}^{h}$, the algorithm assumes it is the fastest client and solves the approximated problem in Eq. \eqref{eq:approximate_problem_formulation}, where the optimization object is $T_{n}(H, \tau_{n}) = \sum_{h'=h}^{H} (\tau_{n} \cdot \mu_{n}^{h'} + \nu_{n}^{h'})$ with $\tau_{n} = \sqrt{\frac{12 F(\mathbf{x}^{h})}{\eta^{2} H L (G^{2} + 18 \sigma^{2})}} $.
Then, the number of rounds $H$ is obtained.
Accordingly, the total completion time for the entire training process $T_{n}$ is calculated according to $H$.
Then, the algorithm selects the fastest client $l$, \ie, $l \leftarrow \mathop{\arg\min}_{n \in \mathcal{N}^{h}} T_{n}$.
However, solving this problem requires information of the entire training process, such as clients' status and network bandwidth.
Unfortunately, since these information are usually time-varying in the dynamic edge system, it is impossible to obtain them in advance.
To this end, we adopt the information in the current round to approximate the unavailable future information.
Therefore, the optimization object is approximated as follows:
\begin{equation} \label{eq:approximated_client_n_completion_time}
    T_{n}(H) = H \cdot ( \sqrt{\frac{12 F(\mathbf{x}^{h})}{\eta^{2} H L (G^{2} + 18 \sigma^{2})}} \cdot \mu_{n}^{h} + \nu_{n}^{h} )
\end{equation}
Notably, there are some variables, such as $L$, $\sigma^{2}$ and $G^{2}$, whose values are unknown at the beginning.
In order to address this issue, when $h = 0$, we adopt an identical predefined local update frequency $\mu_{1}$ for all participating clients, without performing the algorithm.
When $h \geq 1$, each participating client estimates these variables over its local loss function (Lines \ref{alg:client_estimate_begin}-\ref{alg:client_estimate_end} of Alg. \ref{alg:client}) and the PS aggregates them to obtain their specific values (Line \ref{alg:ps_global_aggregation_begin} of Alg. \ref{alg:ps}).

Thirdly, the algorithm determines other clients' local update frequencies based on the fastest client $l$'s completion time in round $h$, \ie, $T_{l}^{h}$ (Lines \ref{alg:ps_determine_other_tau_begin}-\ref{alg:ps_determine_other_tau_end} of Alg. \ref{alg:ps}).
Specifically, for each client $n \in \mathcal{N}^{h}$, the algorithm first derives a frequency space $[\tau_{a}, \tau_{b}]$ according to Eq. \eqref{eq:approximate_waiting_time} and searches the final local update frequency within this space, ensuring the waiting time does not exceed the threshold $\rho$.
Then, $(p_{n}^{h})^{2}$ blocks with the least total update times are selected to form the reduced coefficient $\hat{\mathbf{u}}_{n}^{h}$ for client $n$.
Finally, the algorithm searches the local update frequency $\tau_{n}^{h}$ in $[\tau_{a}, \tau_{b}]$ to minimize the variance among the total update times of all coefficient blocks.

Fourthly, the PS sends the global basis $\mathbf{v}^{h}$, reduced coefficient $\hat{\mathbf{u}}_{n}^{h}$ and local update frequency $\tau_{n}^{h}$ to each participating client $n$ for local training (\ie, Alg. \ref{alg:client}).
Once receiving the updated basis $\bar{\mathbf{v}}_{n}^{h}$ and coefficient $\bar{\mathbf{u}}_{n}^{h}$, as well as the estimated variables' values $L_{n}$, $\sigma^{2}_{n}$ and $G^{2}_{n}$, the PS performs global aggregation (Lines \ref{alg:ps_global_aggregation_begin}-\ref{alg:ps_global_aggregation_end} in Alg. \ref{alg:ps}).
The whole process continues until the total time cost exceeds the budget $T^{max}$.

%% file: content/evaluation.tex
\subsection{Datasets and Models}

\subsubsection{\textbf{Datasets}}
We conduct the experiments over three common datasets: CIFAR-10\cite{krizhevsky2009learning}, ImageNet\cite{russakovsky2015imagenet} and Shakespeare\cite{caldas2018leaf}.
Specifically, CIFAR-10 is an image dataset including 60,000 images (50,000 images for training and 10,000 images for testing), which are 3$\times$32$\times$32 dimensional and evenly from 10 classes.
ImageNet contains 1,281,167 training images, 50,000 validation images and 100,000 test images from 1,000 classes and is more challenging to train the models for visual recognition.
Considering the constrained resource of edge clients, we create a subset of ImageNet, called ImageNet-100, that consists of 100 out of 1,000 classes.
Besides, each image's resolution is resized to 3$\times$144$\times$144.
Shakespeare is a text dataset for next-character prediction built from Shakespeare Dialogues, and includes 422,615 samples with a sequence length of 80.
We split the dataset into 90\% for training and 10\% for testing \cite{caldas2018leaf}.
CIFAR-10 and ImageNet-100 represent the low-resolution and high-resolution computer vision (CV) learning tasks, respectively, while Shakespeare represents the natural language processing (NLP) learning task.

\subsubsection{\textbf{Data Distribution}}
To simulate the non-independent and identically distributed (Non-IID) data, we adopt three different data partition schemes for the three datasets, respectively.
Specifically, we adopt latent Dirichlet allocation (LDA) over CIFAR-10 \cite{jiang2023heterogeneity}, where $\varGamma$\% ($\varGamma =$ 20, 40, 60 and 80) of the samples on each client belong to one class and the remaining samples evenly belong to other classes. 
Particularly, $\varGamma =$ 10 represents the IID setting.
For ImageNet-100, we control that each client lacks $\phi$ ($\phi =$ 20, 40, 60 and 80) classes of samples and the data volume of each class is the same \cite{wang2022accelerating}, where $\phi = $ 0 represents the IID setting.
In our experiments, both $\varGamma$ and $\phi$ are set to 40 by default.
Shakespeare is a natural Non-IID dataset, where the dialogues of each speaking role in each play are regarded as the local data of a specific client \cite{caldas2018leaf} and the Non-IID level is fixed.
For fair comparison, the full test datasets are used to evaluate the models' performance.

\subsubsection{\textbf{Models}}
To validate the universality of the enhanced neural composition technique, we conduct the experiments across several different architectures. 
Firstly, a 4-layer CNN with three 3$\times$3 convolutional layers and one linear output layer is adopted for the CIFAR-10 dataset.
Secondly, we utilize the standard ResNet-18 for the more challenging ImageNet-100 dataset.
Thirdly, for the Shakespeare dataset, we adopt an RNN model and set both the hidden channel size and embedding size to 512 \cite{mei2022resource}.

\subsection{Baselines and Metrics}
\subsubsection{\textbf{Baselines}}
We choose the following four baselines for performance comparison:
\ding{172} \textit{FedAvg} \cite{mcmahan2017communication} transmits and trains the entire models with a fixed (non-dynamic) and identical (non-diverse) local update frequency for all clients.
\ding{173} \textit{ADP} \cite{wang2018edge} dynamically determines the identical local update frequency for all clients in each round on the basis of the constrained resource.
\ding{174} \textit{HeteroFL} \cite{diao2020heterofl} reduces the model width for each client based on its computation power by model pruning.
\ding{175} \textit{Flanc} \cite{mei2022resource} utilizes the neural composition technique to adjust the model width, where the coefficients in different shapes do not share any parameter.

\subsubsection{\textbf{Metrics}}
We employ the following four metrics to evaluate the performance of Heroes and baselines.
\ding{172} \textit{Test accuracy} is measured by the proportion between the amount of the correct samples through model inference and that of all test samples. 
\ding{173} \textit{Average Waiting Time} is calculated by averaging the time each client waits for global aggregation in a round, reflecting the impact of client heterogeneity.
\ding{174} \textit{Completion time} is defined as the total time cost to reach the target accuracy, which reflects the training speed.
\ding{175} \textit{Network Traffic} is the overall size of models (or tensors) transmitted between PS and clients during the training process, which quantifies the communication cost.

\subsection{Experimental Setup}
The experimental environment is built on an AMAX deep learning workstation equipped with an Intel Xeon 5218 CPU, 8 NVIDIA GeForce RTX 3090 GPUs and 256GB RAM.
We simulate an FL system with 100 virtual clients and one PS (each is implemented as a process in the system) on this workstation.
In each round, we randomly activate 10 clients to participate in training.
Specifically, the model training and testing are implemented based on the PyTorch framework\footnote{https://pytorch.org/docs/stable}, and the MPI for Python library\footnote{https://mpi4py.readthedocs.io/en/stable/} is utilized to build up the communication between clients and the PS.

To reflect heterogeneous and dynamic network conditions, we let each client's download speed to fluctuate between 10Mb/s and 20Mb/s \cite{wang2022accelerating}. 
Since the upload speed is usually smaller than the download speed in typical WANs, we configure it to fluctuate between 1Mb/s and 5Mb/s \cite{jiang2023heterogeneity} for each client. 
Besides, considering the clients' computation capabilities are also heterogeneous and dynamic, the time cost of one local iteration on a certain simulated client follows a Gaussian distribution whose mean and variance are derived from the time records on a physical device (\eg, laptop, TX2, Xavier NX, AGX Xavier) \cite{liao2023adaptive}.

\subsection{Evaluation Results}
\subsubsection{\textbf{Training Performance}}
\begin{figure}[t]
	\centering
	\subfigure[CNN on CIFAR-10.]{
		\includegraphics[width=1.6in]{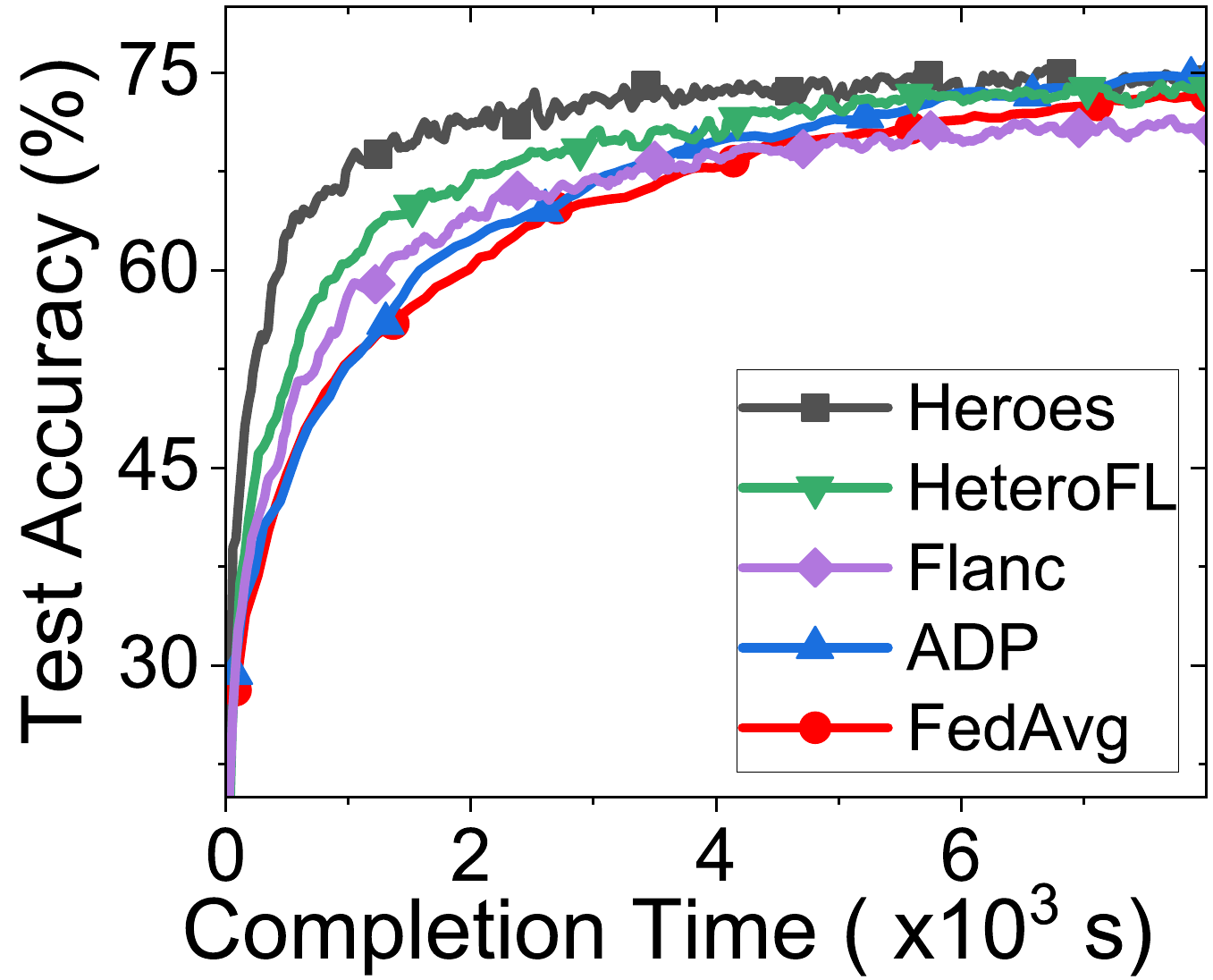}\label{fig:cifar_acc_time}
	}
	\subfigure[ResNet-18 on ImageNet-100.]{
		\includegraphics[width=1.6in]{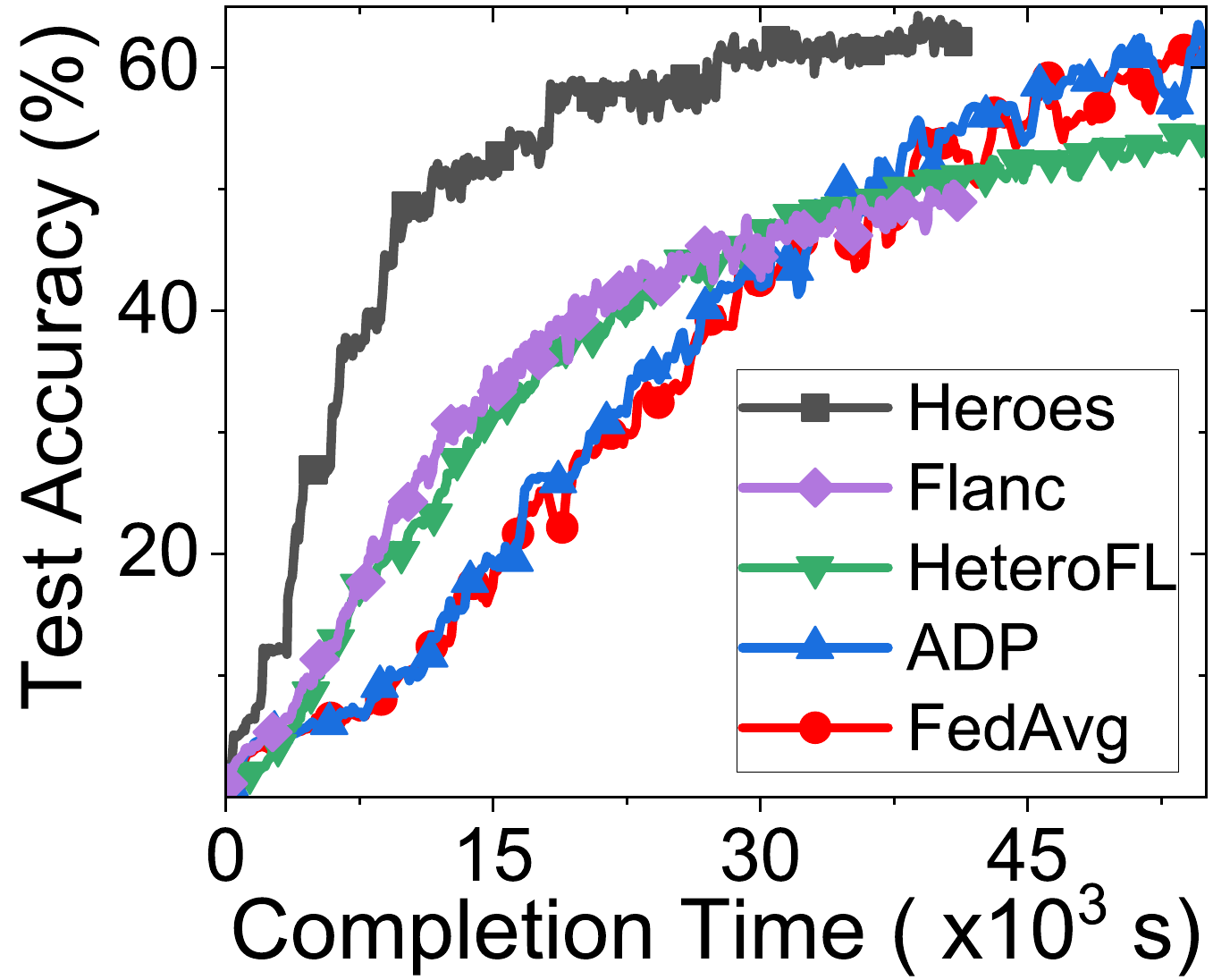}\label{fig:image_acc_time}
	}
    \vspace{-2mm}
	\caption{Training performance of different schemes.}\label{fig:image_dataset_acc_time}
	\vspace{-3mm}
\end{figure}

We implement Heroes and baselines on CIFAR-10 and ImageNet-100 to observe their training performance (\eg, test accuracy).
The results in Fig. \ref{fig:image_dataset_acc_time} show that Heroes converges much faster than the baselines while accomplishing a comparable accuracy.
For instance, by Fig. \ref{fig:cifar_acc_time}, Heroes takes 1,375s to achieve an accuracy of 70\% for CNN on CIFAR-10, while FedAvg, ADP, HeteroFL and Flanc take 4,508s, 3,924s, 3,015s and 3,187s, respectively.
In other words, Heroes can speed up the training process by up to 2.67$\times$ compared to the baselines.
Besides, the model's accuracy in Heroes also surpasses that in the baselines within a given completion time.
For example, Heroes achieves an accuracy of 64.36\% after training ResNet-18 over ImageNet-100 for 40,000s, while that of FedAvg, ADP, HeteroFL and Flanc is 55.22\%, 56.34\%, 52.11\% and 51.89\%, respectively.
In general, within the same time budget, Heroes can improve the test accuracy by about 10.46\% compared with the baselines.
These results demonstrate the advantages of Heroes in accelerating model training.

\subsubsection{\textbf{Impact of Client Heterogeneity}}
\begin{figure}[t]
	\centering
	\subfigure[CNN on CIFAR-10.]{
		\includegraphics[width=1.6in]{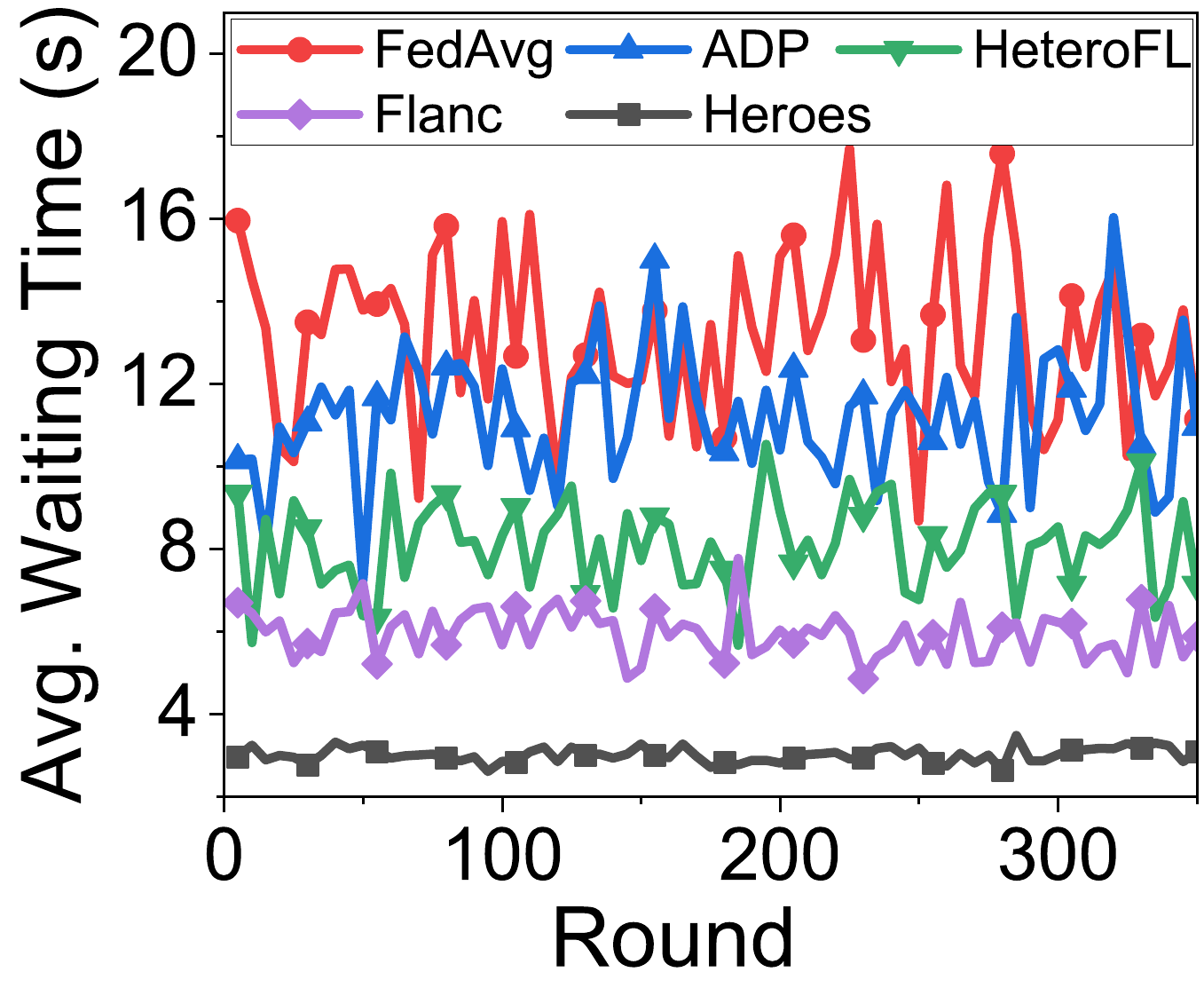}\label{fig:waiting_time_cifar}
	}
	\subfigure[ResNet-18 on ImageNet-100.]{
		\includegraphics[width=1.6in]{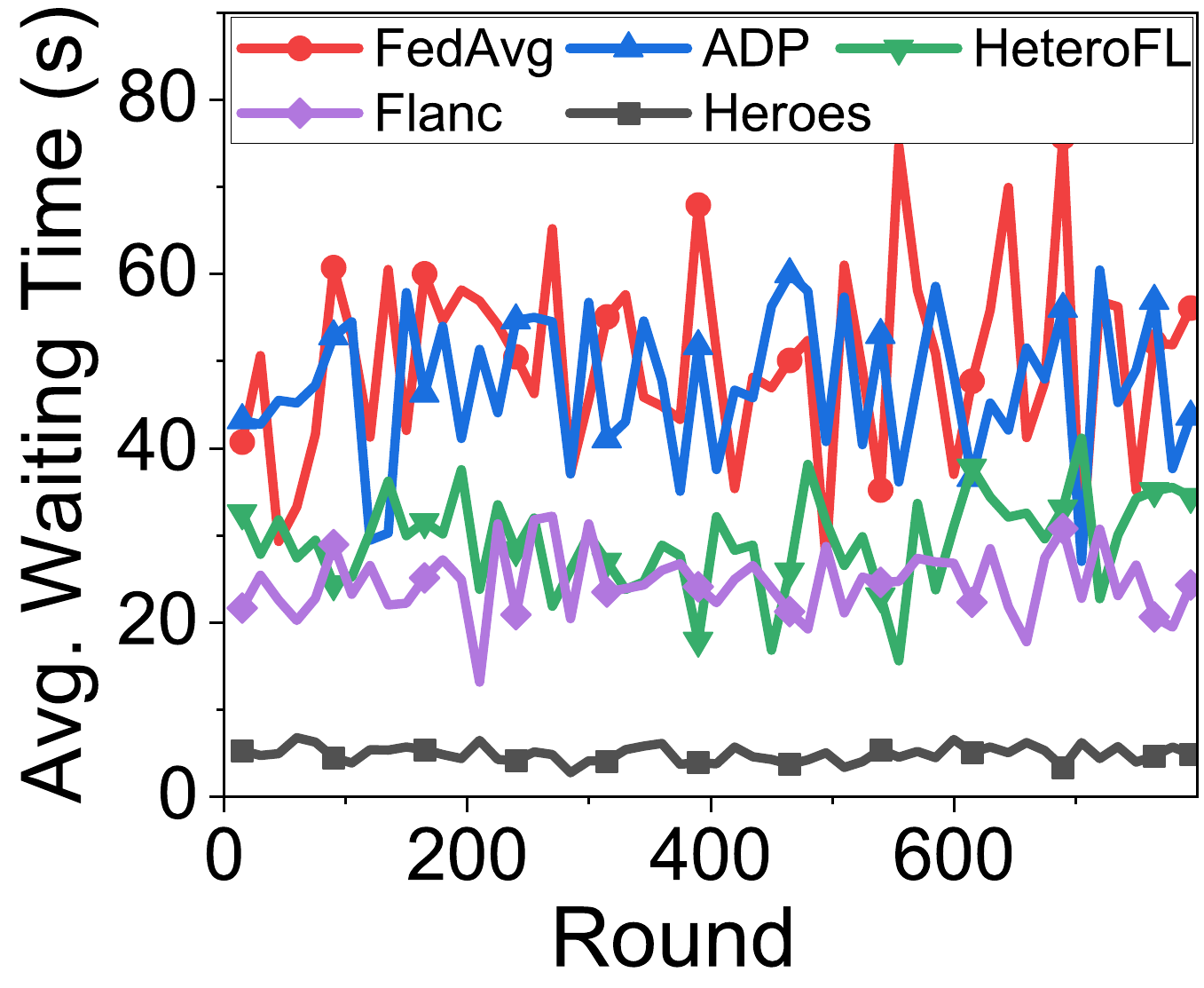}\label{fig:waiting_time_image}
	}
    \vspace{-2mm}
	\caption{Average waiting time of different schemes.}\label{fig:image_dataset_waiting_time}
	\vspace{-3mm}
\end{figure}

To evaluate the impact of client heterogeneity on model training with different schemes, we illustrate the average waiting time each round among the participating clients in Fig. \ref{fig:image_dataset_waiting_time}.
The results show that Heroes incurs much less waiting time than the baselines, which means high robustness against system heterogeneity.
For example, by Fig. \ref{fig:waiting_time_cifar}, the average waiting time in Heroes is 2.86s when training CNN over CIFAR-10, while that in FedAvg, ADP, HeteroFL and Flanc is 15.37s, 11.02s, 8.34s and 5.96s, respectively.
Specifically, both FedAvg and ADP assign the entire model and identical local update frequencies for all clients during each round without considering the system heterogeneity, resulting in non-negligible waiting time.
HeteroFL and Flanc reduce the model width for different clients according to their various computation power.
However, they ignore the heterogeneity in clients' communication capabilities.
For example, the client with a slow upload speed will easily become the straggler and delay the global aggregation.
In addition to reducing the model width, Heroes also adjusts the local update frequencies for different clients to balance their completion time in each round.
Therefore, Heroes can diminish the impact or system heterogeneity significantly.

\subsubsection{\textbf{Resource Consumption}}
\begin{figure}[t]
	\centering
	\subfigure[Traffic Overhead.]{
		\includegraphics[width=1.6in]{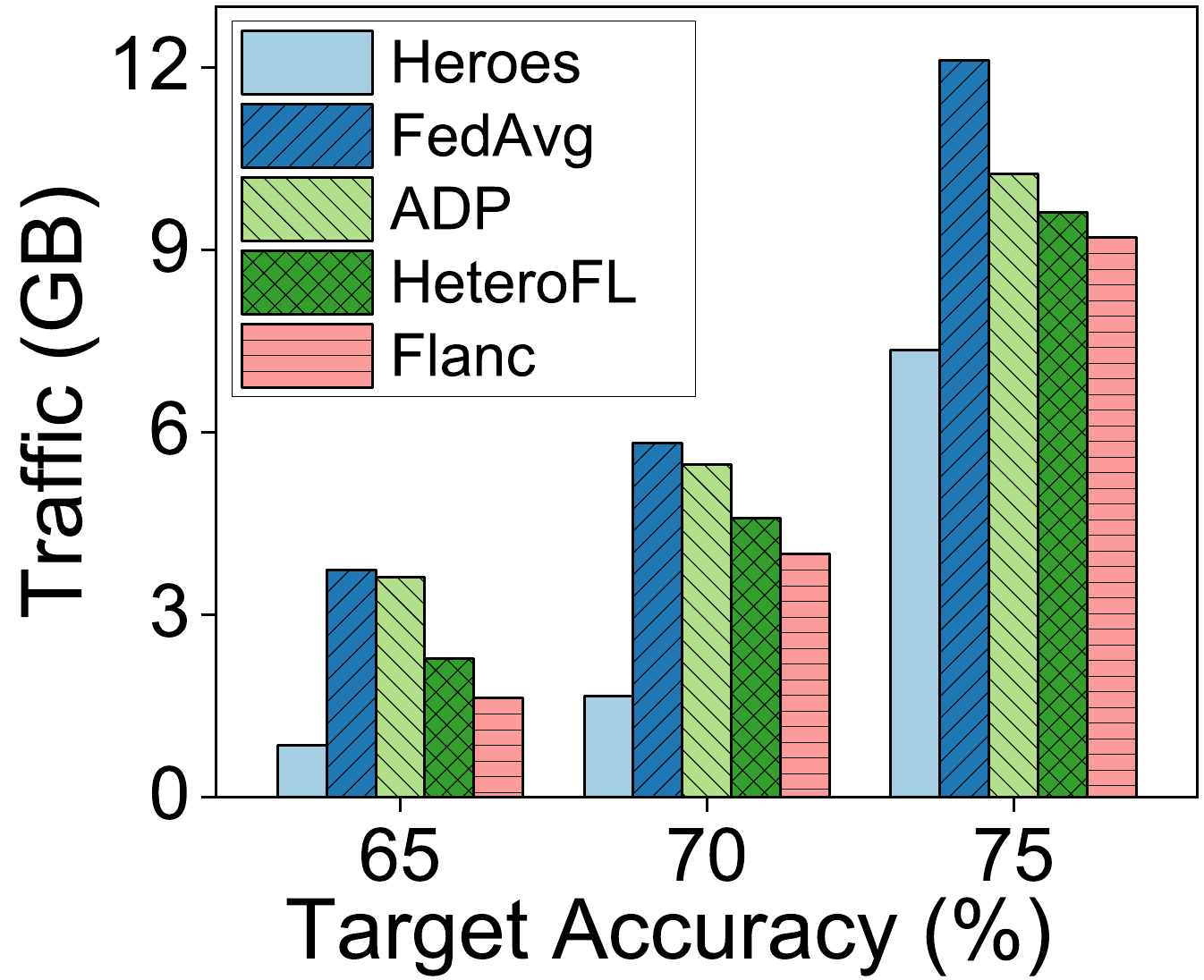}\label{fig:cifar_traffic}
	}
	\subfigure[Completion Time.]{
		\includegraphics[width=1.6in]{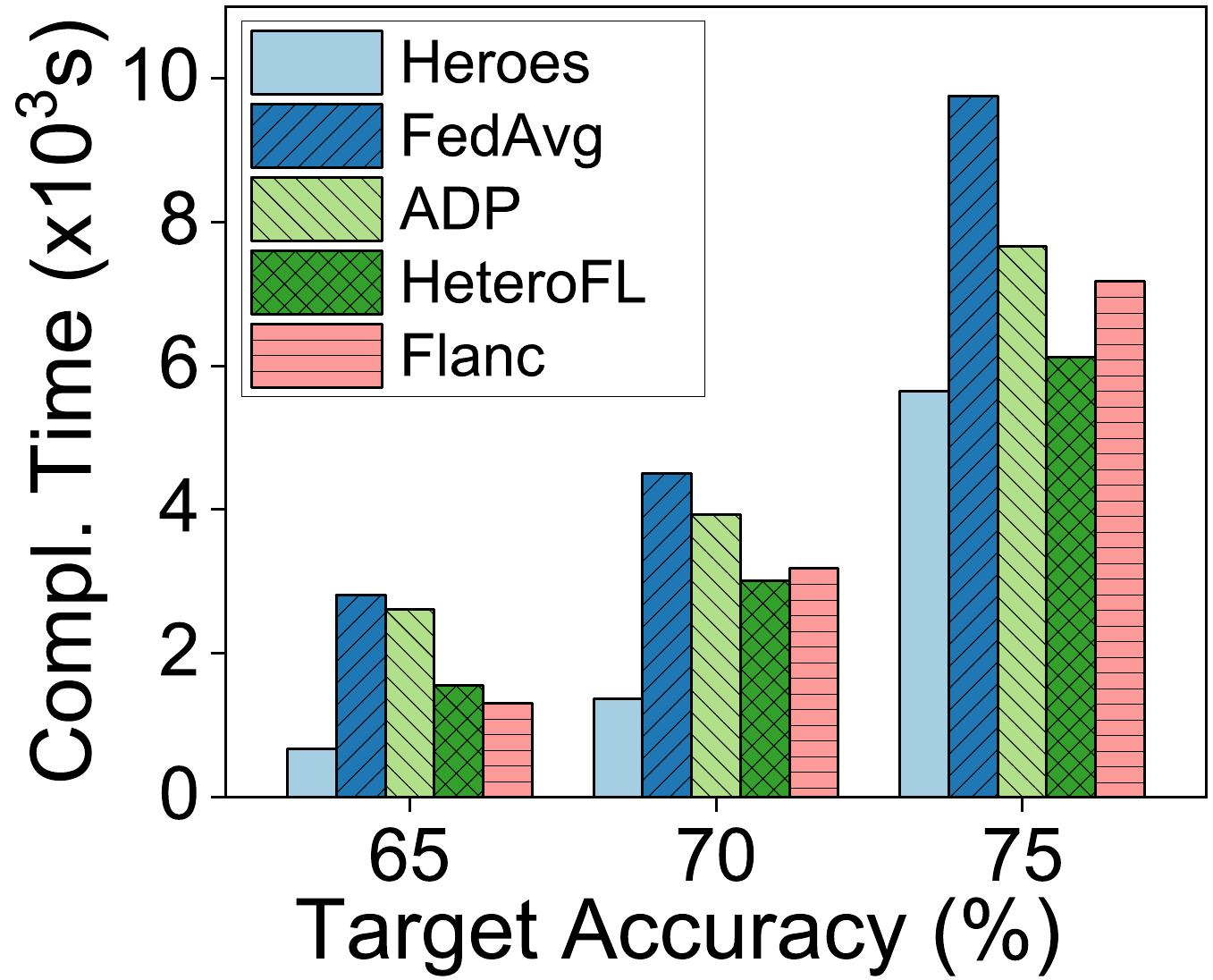}\label{fig:cifar_time}
	}
    \vspace{-2mm}
	\caption{The resources consumption of CNN on CIFAR10.}\label{fig:cifar_resource}
	\vspace{-3mm}
\end{figure}

\begin{figure}[t]
	\centering
	\subfigure[Traffic Overhead.]{
		\includegraphics[width=1.6in]{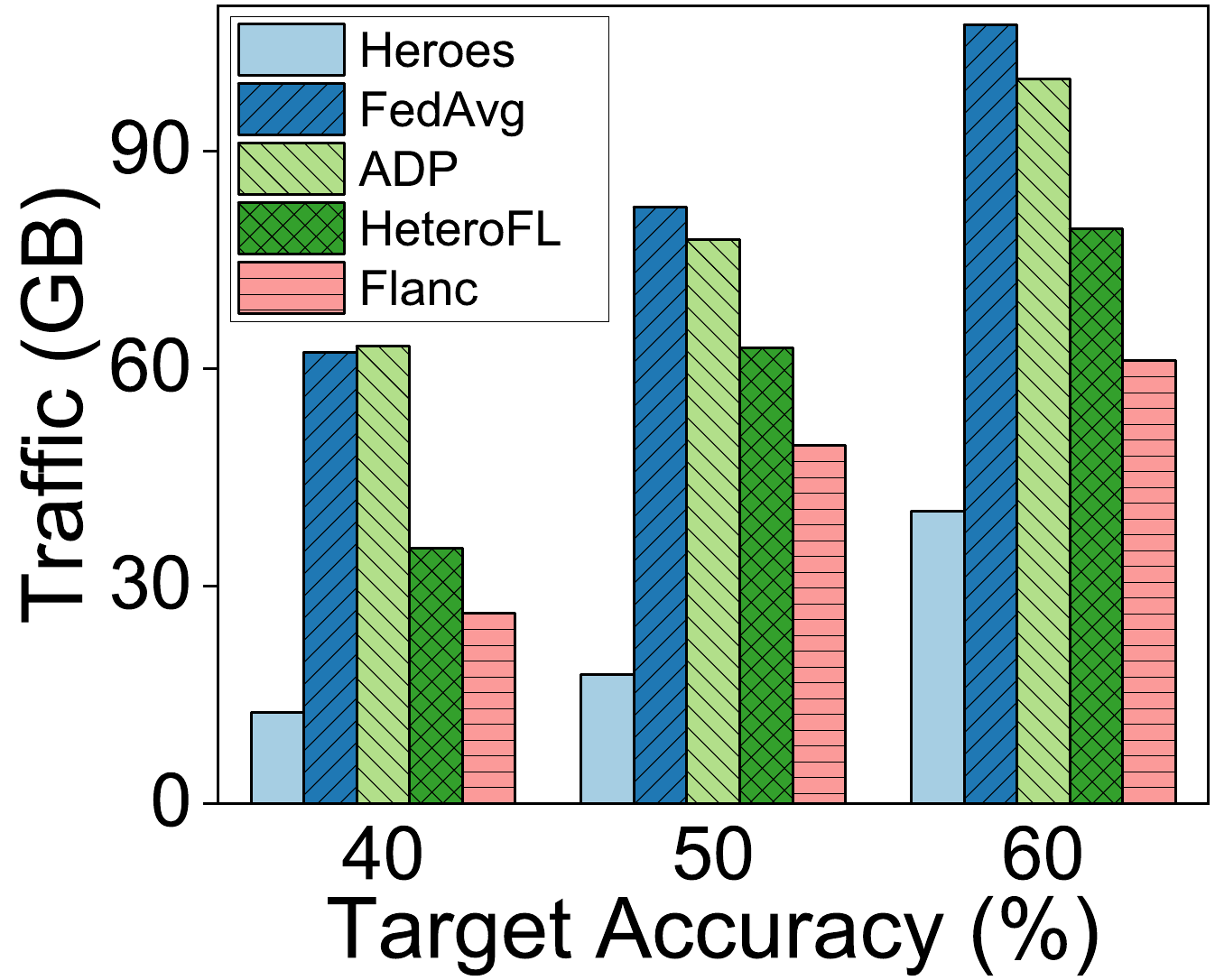}\label{fig:image_traffic}
	}
	\subfigure[Completion Time.]{
		\includegraphics[width=1.6in]{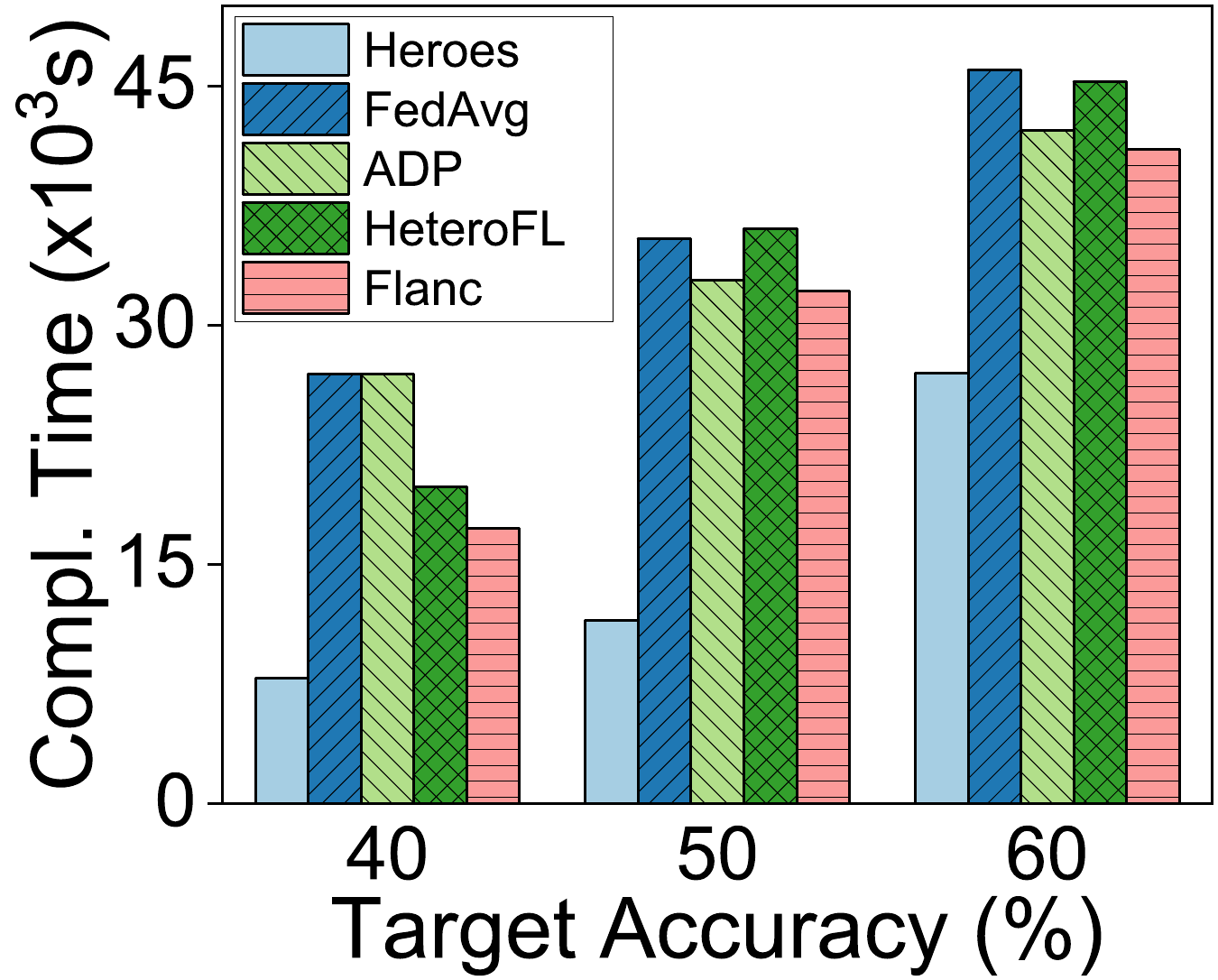}\label{fig:image_time}
	}
    \vspace{-2mm}
	\caption{The resource consumption of ResNet-18 on ImageNet-100.}\label{fig:image_resource}
	\vspace{-3mm}
\end{figure}

We observe the resource consumption (\eg, network traffic and completion time) of five schemes when they achieve different target accuracies on the two image datasets (\eg, 75\% on CIFAR-10, 60\% on ImageNet-100).
The results in Fig. \ref{fig:cifar_resource} and Fig. \ref{fig:image_resource} demonstrate that Heroes can mitigate both the time and traffic costs greatly.
For instance, in Fig. \ref{fig:image_traffic}, to obtain the target accuracy of 50\% on ImageNet-100, the traffic cost of Heroes is 17.81GB, while that of FedAvg, ADP, HeteroFL and Flanc is 82.34GB, 77.75GB, 62.87GB and 49.38GB, respectively.
At the same time, by Fig. \ref{fig:image_time}, Heroes can separately speed up the training process by about 3.09$\times$, 2.86$\times$, 3.15$\times$ and 2.81$\times$, compared with FedAvg, ADP, HeteroFL and Flanc. 
In a word, Heroes achieves the target accuracy fastest with about 2.97$\times$ speedup while reducing the network traffic by about 72.05\% compared with the baselines.

\subsubsection{\textbf{Impact of Non-IID Data}}
\begin{figure}[t]
	\centering
	\subfigure[CNN on CIFAR-10.]{
		\includegraphics[width=1.6in]{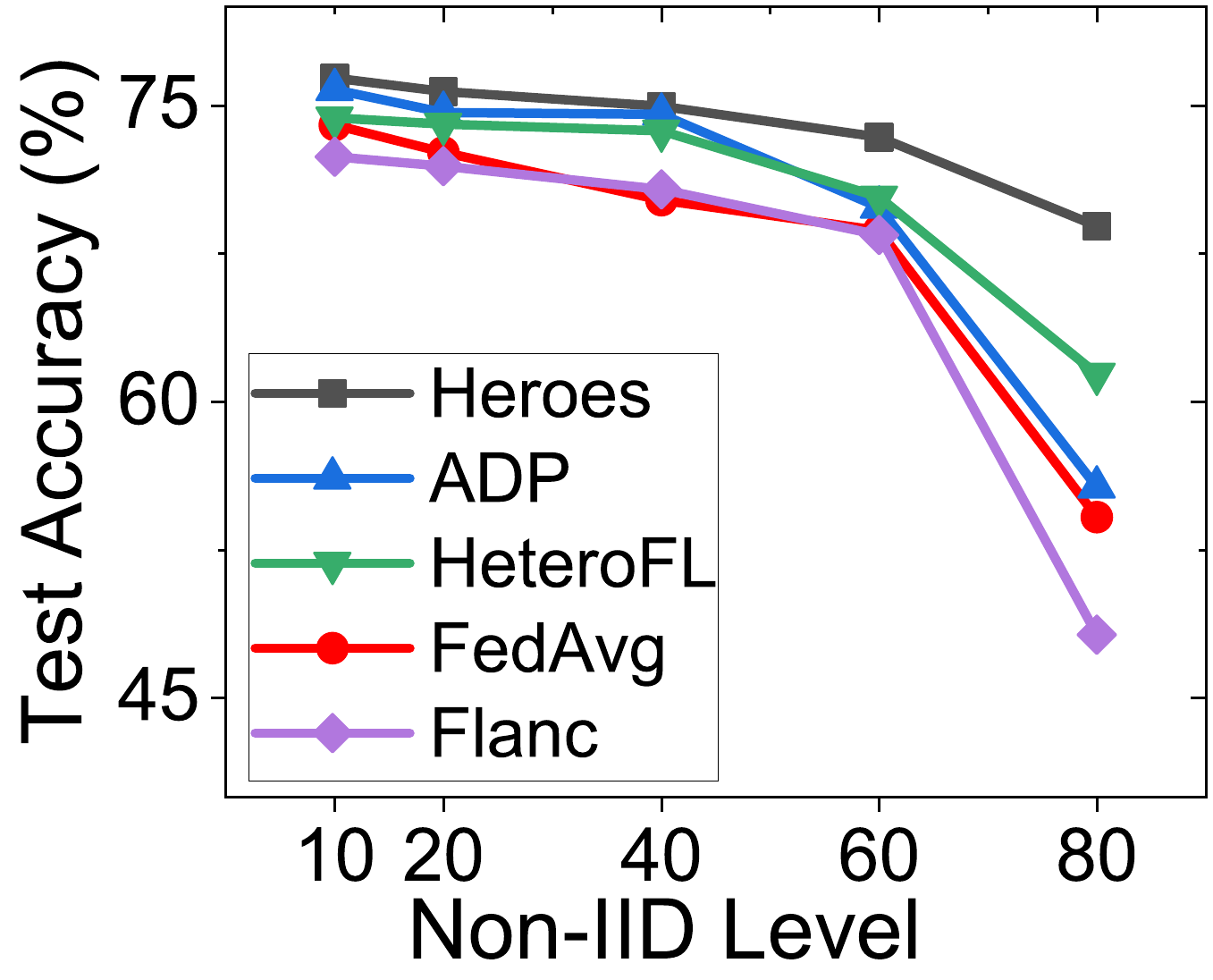}\label{fig:non_iid_level_cifar}
	}
	\subfigure[ResNet-18 on ImageNet-100.]{
		\includegraphics[width=1.6in]{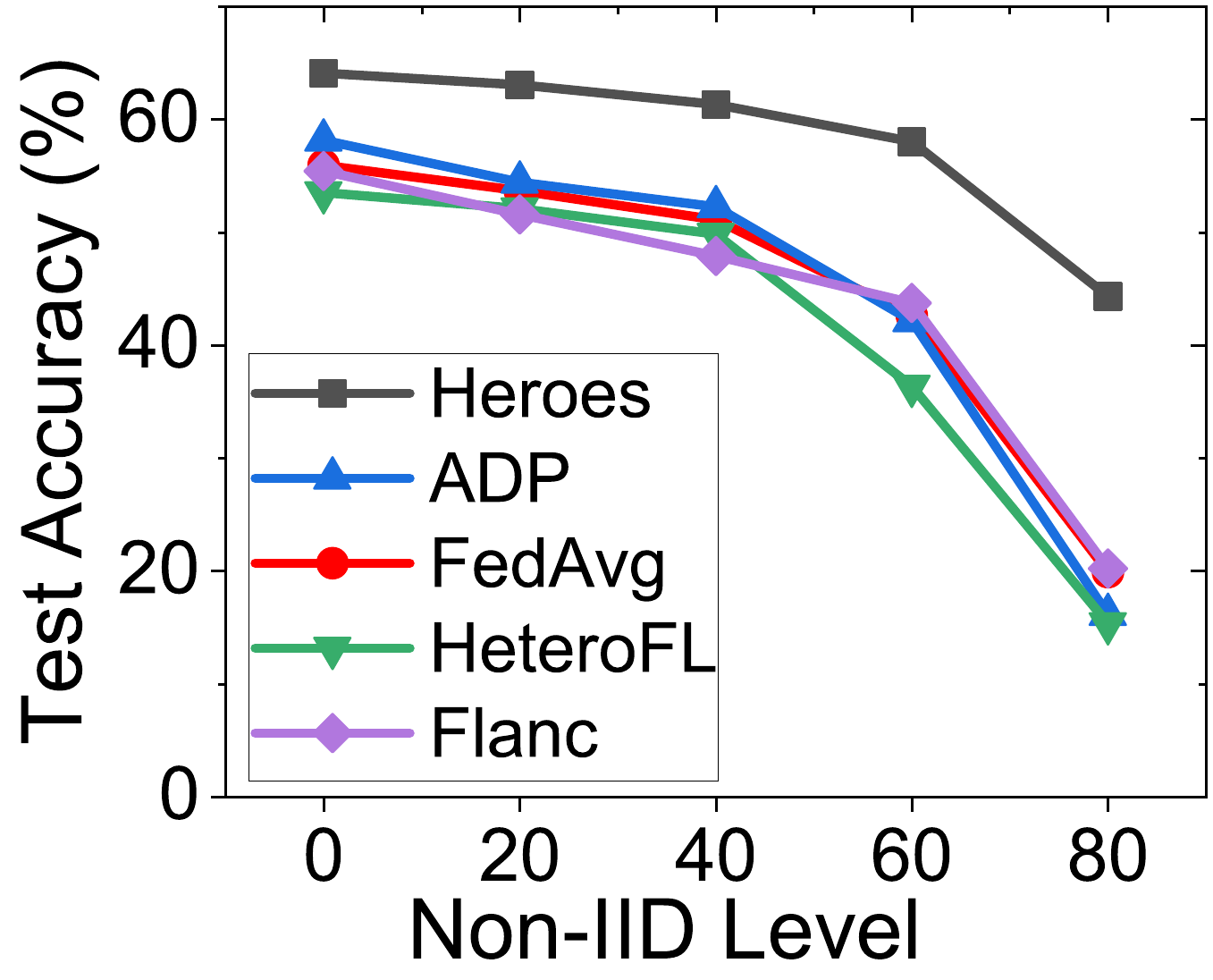}\label{fig:non_iid_level_image}
	}
    \vspace{-2mm}
	\caption{Training performance under different Non-IID levels.}\label{fig:image_dataset_non_iid_level}
	\vspace{-3mm}
\end{figure}

We test these schemes' test accuracies over the CIFAR-10 and ImageNet-100 datasets under different Non-IID levels within a given completion time (800s for CIFAR-10 and 40,000s for ImageNet-100).
The results in Fig. \ref{fig:image_dataset_non_iid_level} indicate that the test accuracy decreases as the Non-IID level increases for all schemes.
For instance, by Fig. \ref{fig:non_iid_level_cifar}, when training CNN on CIFAR-10 with $\varGamma=80$, Heros achieves an accuracy of 68.9\%, which is 14.72\%, 13.21\%, 7.48\% and 24.08\% higher than FedAvg, ADP, HeteroFL and Flanc, respectively.
Heroes investigates the benefits of neural composition technique and adaptive control of local update frequency, which can accelerate the FL process while maintaining the accuracy of the complete model.
Compared to FedAvg and ADP, Heroes will perform more training rounds to achieve higher accuracy within the given time.
Compared to HeteroFL and Flanc, Heroes enables every parameter in the global model to be fully trained over the full range of knowledge, thus eliminating the training bias and achieving better performance.


\begin{figure}[t]
	\centering
	\subfigure[Test Accuracy vs.Time.]{
		\includegraphics[width=1.59in]{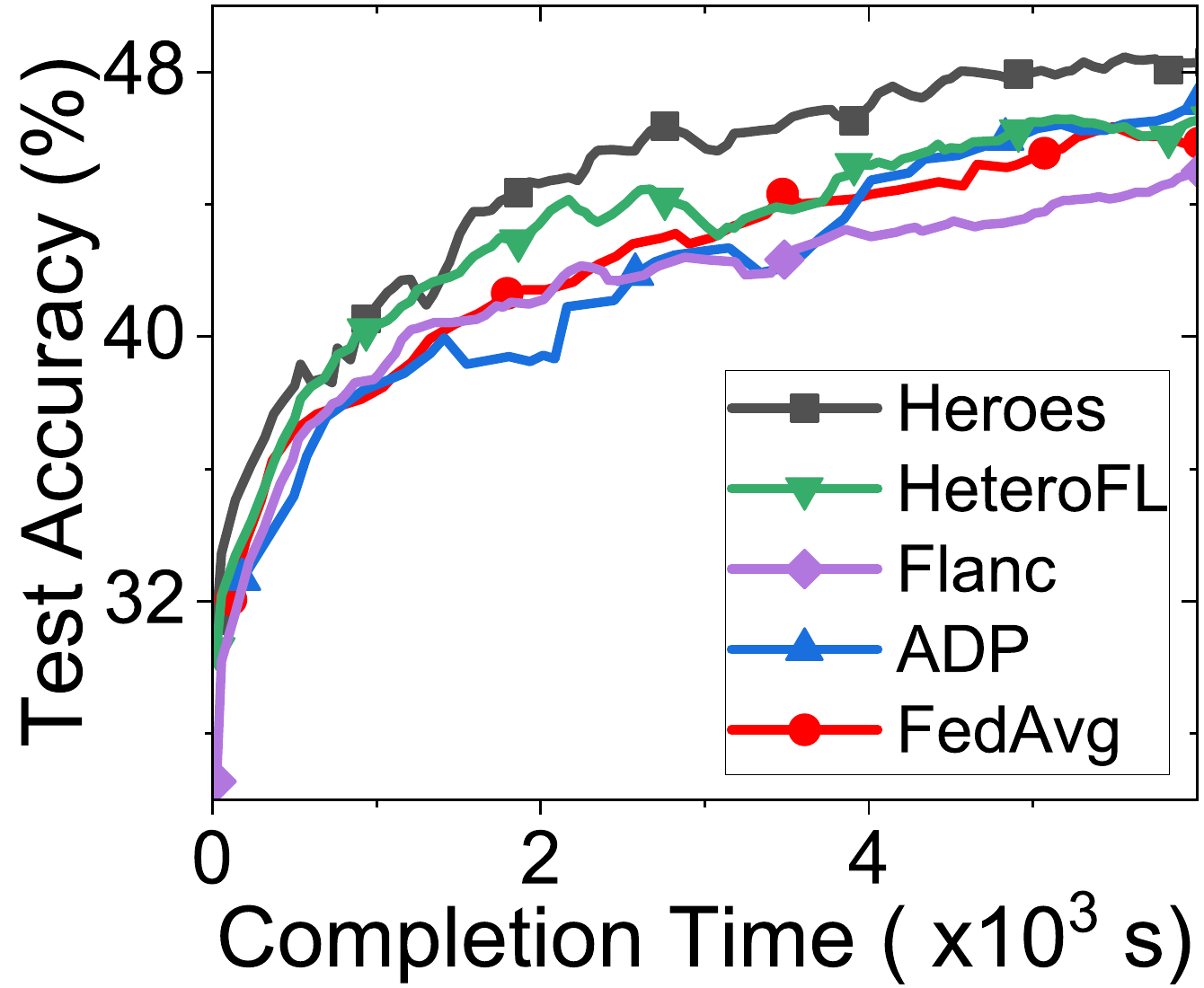}\label{fig:rnn_acc_time}
	}
	\subfigure[Traffic Overhead.]{
		\includegraphics[width=1.59in]{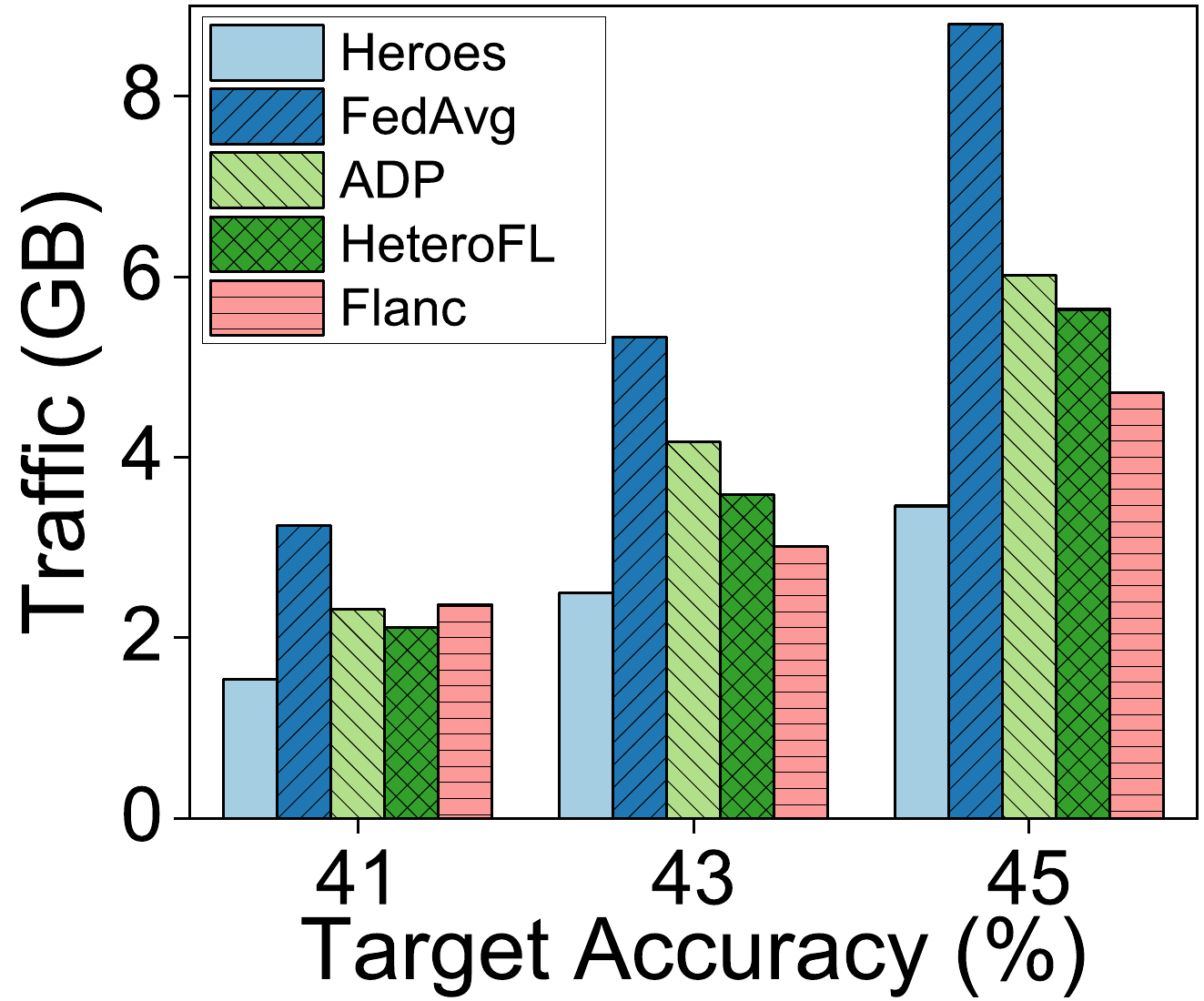}\label{fig:rnn_traffic}
	}
    \vspace{-2mm}
	\caption{The performance of training RNN over Shakespeare}\label{fig:text_dataset_performance}
	\vspace{-3mm}
\end{figure}

\subsubsection{\textbf{Performance on Text Dataset}}
Finally, to verify the generalization of the enhanced neural composition technique, we conduct a set of experiments to train RNN over the text dataset Shakespeare.
The results in Fig. \ref{fig:text_dataset_performance} demonstrate that Heroes can also accomplish better training performance than the baselines on the NLP learning task.
Specifically, according to Fig. \ref{fig:rnn_acc_time}, Heroes takes 1,862s to reach the target accuracy of 45\%, while FedAvg, ADP, HeteroFL and Flanc take 4,183s, 4,015s, 3,182s and 2,874s, respectively.
Besides, by Fig. \ref{fig:rnn_traffic}, compared with FedAvg, ADP, HeteroFL and Flanc, Heroes saves about 60.71\%, 42.57\%, 38.65\% and 26.72\% of network traffic, respectively.
Therefore, compared with baselines, Heroes can provide up to 1.91$\times$ speedup and reduce the traffic consumption by about 45.06\% when training RNN over Shakespeare.


%% file: content/related.tex
As a practical and promising approach, FL has garnered significant interest from both research and industrial communities \cite{kairouz2021advances}.
However, the training efficiency of FL often suffers from resources limitation and client heterogeneity \cite{imteaj2021survey}.
In recent years, many previous works have been proposed to improve the training efficiency for FL.
To tackle the challenge of client heterogeneity, some research \cite{lai2021oort, luo2022tackling} optimizes the client sampling strategy to diminish the heterogeneity degree among participating clients.
Another approach is adjusting the local update frequencies for different clients \cite{li2020federated}, so as to balance their completion time and mitigate the effect of straggler.
However, these approaches have not been able to effectively conserve the limited resources in FL.

Compressing the transmitted gradients is a common way to alleviate the communication overhead \cite{jiang2023heterogeneity, li2021talk, wang2023federated, cui2022optimal, liu2022communication, nori2021fast}.
To further reduce the computation overhead, a natural solution is to prune the global model into a smaller sub-model for training \cite{diao2020heterofl, horvath2021fjord, alam2022fedrolex, li2020lotteryfl, mugunthan2022fedltn, li2021hermes}.
Nevertheless, the compressed or pruned parameters are under-optimized in these approaches, degrading the training performance.
To address this issue, neural composition technique \cite{mei2022resource} is proposed to construct the size-adjustable models using the more efficient low-rank tensors, while enabling every parameter to learn the knowledge from all clients.
However, the global model may get insufficient training, leading to a long completion time, especially in heterogeneous edge networks.

%% file: content/conclusion.tex
In this paper, we have proposed a lightweight FL framework, called Heroes, to address the challenges of resource limitation and client heterogeneity with the enhanced neural composition and adaptive local update.
We have analyzed the convergence of Heroes and designed a greedy-based algorithm to jointly assign proper tensors and local update frequency for each client, which enables every parameter in the global model to benefit from all clients' knowledge and get fully trained.
Extensive experiments demonstrate the effectiveness and advantages of our proposed framework.

%% file: main.bbl
\begin{thebibliography}{10}
\providecommand{\url}[1]{#1}
\csname url@samestyle\endcsname
\providecommand{\newblock}{\relax}
\providecommand{\bibinfo}[2]{#2}
\providecommand{\BIBentrySTDinterwordspacing}{\spaceskip=0pt\relax}
\providecommand{\BIBentryALTinterwordstretchfactor}{4}
\providecommand{\BIBentryALTinterwordspacing}{\spaceskip=\fontdimen2\font plus
\BIBentryALTinterwordstretchfactor\fontdimen3\font minus \fontdimen4\font\relax}
\providecommand{\BIBforeignlanguage}[2]{{%
\expandafter\ifx\csname l@#1\endcsname\relax
\typeout{** WARNING: IEEEtran.bst: No hyphenation pattern has been}%
\typeout{** loaded for the language `#1'. Using the pattern for}%
\typeout{** the default language instead.}%
\else
\language=\csname l@#1\endcsname
\fi
#2}}
\providecommand{\BIBdecl}{\relax}
\BIBdecl

\bibitem{liu2022enhancing}
J.~Liu, Y.~Xu, H.~Xu, Y.~Liao, Z.~Wang, and H.~Huang, ``Enhancing federated learning with intelligent model migration in heterogeneous edge computing,'' in \emph{2022 IEEE 38th International Conference on Data Engineering (ICDE)}.\hskip 1em plus 0.5em minus 0.4em\relax IEEE, 2022, pp. 1586--1597.

\bibitem{mcmahan2017communication}
B.~McMahan, E.~Moore, D.~Ramage, S.~Hampson, and B.~A. y~Arcas, ``Communication-efficient learning of deep networks from decentralized data,'' in \emph{Artificial intelligence and statistics}.\hskip 1em plus 0.5em minus 0.4em\relax PMLR, 2017, pp. 1273--1282.

\bibitem{gao2019auction}
G.~Gao, M.~Xiao, J.~Wu, H.~Huang, S.~Wang, and G.~Chen, ``Auction-based vm allocation for deadline-sensitive tasks in distributed edge cloud,'' \emph{IEEE Transactions on Services Computing}, vol.~14, no.~6, pp. 1702--1716, 2019.

\bibitem{kairouz2021advances}
P.~Kairouz, H.~B. McMahan, B.~Avent, A.~Bellet, M.~Bennis, A.~N. Bhagoji, K.~Bonawitz, Z.~Charles, G.~Cormode, R.~Cummings \emph{et~al.}, ``Advances and open problems in federated learning,'' \emph{Foundations and Trends{\textregistered} in Machine Learning}, vol.~14, no. 1--2, pp. 1--210, 2021.

\bibitem{liu2023finch}
J.~Liu, J.~Yan, H.~Xu, Z.~Wang, J.~Huang, and Y.~Xu, ``Finch: Enhancing federated learning with hierarchical neural architecture search,'' \emph{IEEE Transactions on Mobile Computing}, 2023.

\bibitem{ignatov2019ai}
A.~Ignatov, R.~Timofte, A.~Kulik, S.~Yang, K.~Wang, F.~Baum, M.~Wu, L.~Xu, and L.~Van~Gool, ``Ai benchmark: All about deep learning on smartphones in 2019,'' in \emph{2019 IEEE/CVF International Conference on Computer Vision Workshop (ICCVW)}.\hskip 1em plus 0.5em minus 0.4em\relax IEEE, 2019, pp. 3617--3635.

\bibitem{liu2023yoga}
J.~Liu, J.~Liu, H.~Xu, Y.~Liao, Z.~Wang, and Q.~Ma, ``Yoga: Adaptive layer-wise model aggregation for decentralized federated learning,'' \emph{IEEE/ACM Transactions on Networking}, 2023.

\bibitem{wang2021resource}
Z.~Wang, H.~Xu, J.~Liu, H.~Huang, C.~Qiao, and Y.~Zhao, ``Resource-efficient federated learning with hierarchical aggregation in edge computing,'' in \emph{IEEE INFOCOM 2021-IEEE Conference on Computer Communications}.\hskip 1em plus 0.5em minus 0.4em\relax IEEE, 2021, pp. 1--10.

\bibitem{he2016deep}
K.~He, X.~Zhang, S.~Ren, and J.~Sun, ``Deep residual learning for image recognition,'' in \emph{Proceedings of the IEEE conference on computer vision and pattern recognition}, 2016, pp. 770--778.

\bibitem{jiang2022fedmp}
Z.~Jiang, Y.~Xu, H.~Xu, Z.~Wang, C.~Qiao, and Y.~Zhao, ``Fedmp: Federated learning through adaptive model pruning in heterogeneous edge computing,'' in \emph{2022 IEEE 38th International Conference on Data Engineering (ICDE)}.\hskip 1em plus 0.5em minus 0.4em\relax IEEE, 2022, pp. 767--779.

\bibitem{li2021talk}
L.~Li, D.~Shi, R.~Hou, H.~Li, M.~Pan, and Z.~Han, ``To talk or to work: Flexible communication compression for energy efficient federated learning over heterogeneous mobile edge devices,'' in \emph{IEEE INFOCOM 2021-IEEE Conference on Computer Communications}.\hskip 1em plus 0.5em minus 0.4em\relax IEEE, 2021, pp. 1--10.

\bibitem{liu2022communication}
H.~Liu, F.~He, and G.~Cao, ``Communication-efficient federated learning for heterogeneous edge devices based on adaptive gradient quantization,'' \emph{arXiv preprint arXiv:2212.08272}, 2022.

\bibitem{diao2020heterofl}
E.~Diao, J.~Ding, and V.~Tarokh, ``Heterofl: Computation and communication efficient federated learning for heterogeneous clients,'' in \emph{International Conference on Learning Representations}, 2020.

\bibitem{horvath2021fjord}
S.~Horvath, S.~Laskaridis, M.~Almeida, I.~Leontiadis, S.~Venieris, and N.~Lane, ``Fjord: Fair and accurate federated learning under heterogeneous targets with ordered dropout,'' \emph{Advances in Neural Information Processing Systems}, vol.~34, pp. 12\,876--12\,889, 2021.

\bibitem{mei2022resource}
Y.~Mei, P.~Guo, M.~Zhou, and V.~Patel, ``Resource-adaptive federated learning with all-in-one neural composition,'' in \emph{Advances in Neural Information Processing Systems}, 2022.

\bibitem{li2022pyramidfl}
C.~Li, X.~Zeng, M.~Zhang, and Z.~Cao, ``Pyramidfl: A fine-grained client selection framework for efficient federated learning,'' in \emph{Proceedings of the 28th Annual International Conference on Mobile Computing And Networking}, 2022, pp. 158--171.

\bibitem{yu2019parallel}
H.~Yu, S.~Yang, and S.~Zhu, ``Parallel restarted sgd with faster convergence and less communication: Demystifying why model averaging works for deep learning,'' in \emph{Proceedings of the AAAI Conference on Artificial Intelligence}, vol.~33, no.~01, 2019, pp. 5693--5700.

\bibitem{phan2020stable}
A.-H. Phan, K.~Sobolev, K.~Sozykin, D.~Ermilov, J.~Gusak, P.~Tichavsk{\`y}, V.~Glukhov, I.~Oseledets, and A.~Cichocki, ``Stable low-rank tensor decomposition for compression of convolutional neural network,'' in \emph{Computer Vision--ECCV 2020: 16th European Conference, Glasgow, UK, August 23--28, 2020, Proceedings, Part XXIX 16}.\hskip 1em plus 0.5em minus 0.4em\relax Springer, 2020, pp. 522--539.

\bibitem{zou2019improved}
D.~Zou and Q.~Gu, ``An improved analysis of training over-parameterized deep neural networks,'' \emph{Advances in neural information processing systems}, vol.~32, 2019.

\bibitem{russakovsky2015imagenet}
O.~Russakovsky, J.~Deng, H.~Su, J.~Krause, S.~Satheesh, S.~Ma, Z.~Huang, A.~Karpathy, A.~Khosla, M.~Bernstein \emph{et~al.}, ``Imagenet large scale visual recognition challenge,'' \emph{International journal of computer vision}, vol. 115, pp. 211--252, 2015.

\bibitem{liu2023adaptive}
J.~Liu, Q.~Zeng, H.~Xu, Y.~Xu, Z.~Wang, and H.~Huang, ``Adaptive block-wise regularization and knowledge distillation for enhancing federated learning,'' \emph{IEEE/ACM Transactions on Networking}, 2023.

\bibitem{wang2020towards}
C.~Wang, Y.~Yang, and P.~Zhou, ``Towards efficient scheduling of federated mobile devices under computational and statistical heterogeneity,'' \emph{IEEE Transactions on Parallel and Distributed Systems}, vol.~32, no.~2, pp. 394--410, 2020.

\bibitem{li2020federated}
T.~Li, A.~K. Sahu, M.~Zaheer, M.~Sanjabi, A.~Talwalkar, and V.~Smith, ``Federated optimization in heterogeneous networks,'' \emph{Proceedings of Machine learning and systems}, vol.~2, pp. 429--450, 2020.

\bibitem{xu2022adaptive}
Y.~Xu, Y.~Liao, H.~Xu, Z.~Ma, L.~Wang, and J.~Liu, ``Adaptive control of local updating and model compression for efficient federated learning,'' \emph{IEEE Transactions on Mobile Computing}, 2022.

\bibitem{jiang2023computation}
Z.~Jiang, Y.~Xu, H.~Xu, Z.~Wang, J.~Liu, Q.~Chen, and C.~Qiao, ``Computation and communication efficient federated learning with adaptive model pruning,'' \emph{IEEE Transactions on Mobile Computing}, 2023.

\bibitem{zhan2020incentive}
Y.~Zhan and J.~Zhang, ``An incentive mechanism design for efficient edge learning by deep reinforcement learning approach,'' in \emph{IEEE INFOCOM 2020-IEEE conference on computer communications}.\hskip 1em plus 0.5em minus 0.4em\relax IEEE, 2020, pp. 2489--2498.

\bibitem{krizhevsky2009learning}
A.~Krizhevsky, G.~Hinton \emph{et~al.}, ``Learning multiple layers of features from tiny images,'' 2009.

\bibitem{caldas2018leaf}
S.~Caldas, S.~M.~K. Duddu, P.~Wu, T.~Li, J.~Kone{\v{c}}n{\`y}, H.~B. McMahan, V.~Smith, and A.~Talwalkar, ``Leaf: A benchmark for federated settings,'' \emph{arXiv preprint arXiv:1812.01097}, 2018.

\bibitem{jiang2023heterogeneity}
Z.~Jiang, Y.~Xu, H.~Xu, Z.~Wang, and C.~Qian, ``Heterogeneity-aware federated learning with adaptive client selection and gradient compression,'' in \emph{IEEE INFOCOM 2023-IEEE Conference on Computer Communications}.\hskip 1em plus 0.5em minus 0.4em\relax IEEE, 2023, pp. 1--10.

\bibitem{wang2022accelerating}
L.~Wang, Y.~Xu, H.~Xu, M.~Chen, and L.~Huang, ``Accelerating decentralized federated learning in heterogeneous edge computing,'' \emph{IEEE Transactions on Mobile Computing}, 2022.

\bibitem{wang2018edge}
S.~Wang, T.~Tuor, T.~Salonidis, K.~K. Leung, C.~Makaya, T.~He, and K.~Chan, ``When edge meets learning: Adaptive control for resource-constrained distributed machine learning,'' in \emph{IEEE INFOCOM 2018-IEEE conference on computer communications}.\hskip 1em plus 0.5em minus 0.4em\relax IEEE, 2018, pp. 63--71.

\bibitem{liao2023adaptive}
Y.~Liao, Y.~Xu, H.~Xu, L.~Wang, and C.~Qian, ``Adaptive configuration for heterogeneous participants in decentralized federated learning,'' in \emph{IEEE INFOCOM 2023-IEEE Conference on Computer Communications}.\hskip 1em plus 0.5em minus 0.4em\relax IEEE, 2023, pp. 1--10.

\bibitem{imteaj2021survey}
A.~Imteaj, U.~Thakker, S.~Wang, J.~Li, and M.~H. Amini, ``A survey on federated learning for resource-constrained iot devices,'' \emph{IEEE Internet of Things Journal}, vol.~9, no.~1, pp. 1--24, 2021.

\bibitem{lai2021oort}
F.~Lai, X.~Zhu, H.~V. Madhyastha, and M.~Chowdhury, ``Oort: Efficient federated learning via guided participant selection,'' in \emph{15th $\{$USENIX$\}$ Symposium on Operating Systems Design and Implementation ($\{$OSDI$\}$ 21)}, 2021, pp. 19--35.

\bibitem{luo2022tackling}
B.~Luo, W.~Xiao, S.~Wang, J.~Huang, and L.~Tassiulas, ``Tackling system and statistical heterogeneity for federated learning with adaptive client sampling,'' in \emph{IEEE INFOCOM 2022-IEEE conference on computer communications}.\hskip 1em plus 0.5em minus 0.4em\relax IEEE, 2022, pp. 1739--1748.

\bibitem{wang2023federated}
S.~Wang, J.~Perazzone, M.~Ji, and K.~S. Chan, ``Federated learning with flexible control,'' in \emph{IEEE INFOCOM 2023-IEEE Conference on Computer Communications}.\hskip 1em plus 0.5em minus 0.4em\relax IEEE, 2023, pp. 1--10.

\bibitem{cui2022optimal}
L.~Cui, X.~Su, Y.~Zhou, and J.~Liu, ``Optimal rate adaption in federated learning with compressed communications,'' in \emph{IEEE INFOCOM 2022-IEEE Conference on Computer Communications}.\hskip 1em plus 0.5em minus 0.4em\relax IEEE, 2022, pp. 1459--1468.

\bibitem{nori2021fast}
M.~K. Nori, S.~Yun, and I.-M. Kim, ``Fast federated learning by balancing communication trade-offs,'' \emph{IEEE Transactions on Communications}, vol.~69, no.~8, pp. 5168--5182, 2021.

\bibitem{alam2022fedrolex}
S.~Alam, L.~Liu, M.~Yan, and M.~Zhang, ``Fedrolex: Model-heterogeneous federated learning with rolling sub-model extraction,'' \emph{Advances in Neural Information Processing Systems}, vol.~35, pp. 29\,677--29\,690, 2022.

\bibitem{li2020lotteryfl}
A.~Li, J.~Sun, B.~Wang, L.~Duan, S.~Li, Y.~Chen, and H.~Li, ``Lotteryfl: Personalized and communication-efficient federated learning with lottery ticket hypothesis on non-iid datasets,'' \emph{arXiv preprint arXiv:2008.03371}, 2020.

\bibitem{mugunthan2022fedltn}
V.~Mugunthan, E.~Lin, V.~Gokul, C.~Lau, L.~Kagal, and S.~Pieper, ``Fedltn: Federated learning for sparse and personalized lottery ticket networks,'' in \emph{European Conference on Computer Vision}.\hskip 1em plus 0.5em minus 0.4em\relax Springer, 2022, pp. 69--85.

\bibitem{li2021hermes}
A.~Li, J.~Sun, P.~Li, Y.~Pu, H.~Li, and Y.~Chen, ``Hermes: an efficient federated learning framework for heterogeneous mobile clients,'' in \emph{Proceedings of the 27th Annual International Conference on Mobile Computing and Networking}, 2021, pp. 420--437.

\end{thebibliography}
